\documentclass[preprint,aps,epsfig,floatfix]{revtex4}

\usepackage{graphicx}
\usepackage{dcolumn}
\usepackage{bm}
\usepackage{amsmath}
\begin{document}

\title{Single-Particle Density of States of a Superconductor with a
  Spatially Varying Gap and Phase Fluctuations}

\author{Daniel Valdez-Balderas} \email{balderas@mps.ohio-state.edu}

\author{David Stroud} \email{stroud@mps.ohio-state.edu}


\affiliation{Department of Physics, The Ohio State University, Columbus,
  Ohio 43210}

\date{\today}

\begin{abstract}
  Recent experiments have shown that the superconducting energy gap in
  some cuprates is spatially inhomogeneous.  Motivated by these
  experiments, and using exact diagonalization of a model $d$-wave
  Hamiltonian, combined with Monte Carlo simulations of a
  Ginzburg-Landau free energy functional, we have calculated the
  single-particle density of states LDOS$(\omega,r)$ of a model
  high-T$_c$ superconductor as a 
  function of temperature.  Our calculations include both quenched
  disorder in the pairing potential and thermal fluctuations in both
  phase and amplitude of the superconducting gap.  Most of our
  calculations assume two types of superconducting regions: $\alpha$,
  with a small gap and large superfluid density, and $\beta$, with the
  opposite.  If the $\beta$ regions are randomly embedded in an
  $\alpha$ host, the LDOS on the $\alpha$ sites still has a sharp
  coherence peak at $T = 0$, but the $\beta$ component does not, in
  agreement with experiment.  An ordered arrangement of $\beta$
  regions leads to oscillations in the LDOS as a function of energy.
  The model leads to a superconducting transition temperature $T_c$
  well below the pseudogap temperature $T_{c0}$, and has a spatially
  varying gap at very low $T$, both consistent with experiments in
  underdoped Bi2212.  Our calculated LDOS$(\omega,r)$ shows coherence
  peaks for $T < T_c$, which disappear for $T > T_c$, in agreement
  with previous work considering phase but not amplitude fluctuations
  in a homogeneous superconductor.  Well above $T_c$, the gap in the
  LDOS disappears.

\end{abstract}

\maketitle
\section{Introduction}

According to low temperature scanning tunneling microscopy (STM)
experiments,
the local density of states (LDOS) of some cuprate materials have
spatial variations \cite{cren, pan_davis, howald1, lang_davis,
  howald2, fang, kato_sakata, mashima}.  Among the cuprates,
Bi$_{2}$Sr$_{2}$CaCu$_{2}$O$_{8+x}$ (Bi2212) is one of the most
extensively studied in STM experiments. The LDOS spectrum shows that
some regions of that material, which we will call $\alpha$-regions,
have a small energy gap with large and narrow coherence peaks
(reminiscent of the spectra observed in bulk superconducting
materials), while other regions, which we will call $\beta$-regions,
have a larger gap, but smaller and broadened peaks (which are
reminiscent of the spectra seen in bulk pseudogap phase of some
materials . These inhomogeneities occur on length scales of order
30$\AA$. Because at low doping concentrations $\alpha$ regions with
``good'' superconductivity are immersed in more metallic or
semiconducting $\beta$ regions, some workers have made an analogy
between these materials and granular superconductors
\cite{lang_davis, joglekar}: superconducting domains spatially
separated from one another by non-superconducting regions, but
connected through proximity effect or Josephson
tunneling.


At present there is no general agreement regarding the origin of the
inhomogeneities in the cuprate superconductors ---whether they are in
charge density, spin density, LDOS, or other properties
\cite{scalapino_nunner_hirschfeld}.  One hypothesis is that these
inhomogeneities originate in a process of self organization due to
competing orders \cite{zaanen, emery_lin, emery_kivelson, low_emery,
  emery_kivelson_tranquada, jamei, valdez_stroud}.  In another
approach, the spatially varying properties of the cuprates are
attributed to crystal defects or impurities.  In particular, it has
been suggested that the inhomogeneities in the LDOS originate in the
random spatial distribution of dopant atoms near the copper oxide
(CuO$_2$) planes \cite{pan_davis, martin_balatsky, wang_pan, atkinson,
  nunner_andersen_hirscfeld}.


Several workers have studied the LDOS of inhomogeneous
superconductors at low $T$.  For example, Ghosal {\it et
al}\cite{ghosal} have calculated the LDOS of a strongly disordered
$s$-wave superconducting layer in two dimensions, solving the
Bogoliubov-de Gennes equations self-consistently.  They have also
done similar work on a model of $d$-wave
superconductivity\cite{ghosal1}.
%
%
Fang {\it et al.}\cite{fang}, using a Green's function approach,
computed the zero temperature LDOS of a model lattice Hamiltonian in
which one small region of the lattice has an different (either
suppressed or enhanced) pairing strength than the rest; they find
good agreement with experiments.  Cheng and Su \cite{cheng_su} have
also explored how the LDOS is affected by a single spatial
inhomogeneity in the pairing strength of a BCS Hamiltonian; they
find that an inhomogeneity with an LDOS most closely resembling the
experimental results is produced by an inhomogeneity with a
cone-shaped distribution of the pairing strength; this work thus
suggests that it is the small-length-scale variation of the pairing
strength that causes incoherence in the LDOS.
Mayr {\it et al.} \cite{mayr_dagotto} have studied a phenomenological
model with quenched disorder and observed a pseudogap in the LDOS
caused by a mixture of antiferromagnetism and superconductivity, while
Jamei {\it et al.} \cite{jamei_kapitulnik_kivelson} have investigated
the low order moments of the LDOS and their relation to the local form
of the Hamiltonian.

In this paper we propose a phenomenological approach to study the
effect of inhomogeneities on the LDOS in a model for cuprate
superconductors. The model is a mean-field BCS Hamiltonian with
$d$-wave symmetry, in which the pairing-field is inhomogeneous and
also undergoes thermal fluctuations in both phase and amplitude at finite
temperatures $T$.
It has been argued \cite{scalapino_nunner_hirschfeld} that the
superconducting state of optimally doped to overdoped cuprates is
well described by the BCS theory which includes a $d$-wave gap and
scattering from defects outside the Cu0$_2$ plane. Instead of
including such defects explicitly in our BCS Hamiltonian, we
implicitly include their possible effects through inhomogeneities of
the pairing-field amplitude.
Furthermore, instead of self-consistently solving the Bogoliubov-de
Gennes equations resulting from this model, we obtain the magnitude
and phase of the complex pairing-field from Monte Carlo simulations
based on a Ginzburg Landau free energy functional.   Thus the
procedure is as follows.  First, we set the parameters of the
Ginzburg Landau free energy functional from experiments.  Next,
using Monte Carlo simulations of this free energy, we obtain the
pairing-field amplitudes which we then include in the BCS Hamiltonian.
Finally, we diagonalize the latter in order to obtain the LDOS.

Now in optimally or nearly optimally doped Bi2212, the layers consist of
randomly distributed $\beta$-regions immersed in a
majority background of $\alpha$-regions \cite{pan_davis}.
We therefore choose
Ginzburg-Landau parameters so as to reproduce this morphology at $T
= 0$, then carry out simulations at both zero and finite $T$ to
obtain the LDOS in the different spatial regions.

At $T = 0$ we compare
these simulation results to those obtained using ordered instead of
random arrangements of inhomogeneities.
We find that the LDOS of the random systems much more closely
resemble experiment.  Specifically, regions with a small gap have
sharp coherence peaks, while large-gap regions show lower and
broader peaks.   By contrast, systems with ordered inhomogeneities
have LDOS spectra with sharp coherence peaks which oscillate as a
function of energy.
In the ordered systems, the coherence peaks in the small-gap regions
strongly resemble those observed in a homogeneous small-gap system.
But the spectral peaks in the large-gap regions dramatically differ
from those in the corresponding homogeneous and disordered cases.

Because the spectra of disordered systems more closely resemble
experiments, we have also studied the evolution of the LDOS in these
systems with increasing $T$.  We consider both $T < T_c$ and $T >
T_c$, where $T_c$ is the phase-ordering transition temperature
(equivalent to the Kosterlitz-Thouless transition temperature for this
two-dimensional system).  In both the $\alpha$ and $\beta$ regions, we
find that the spectral gap starts to fill in as $T$ increases, and the
spectral peaks broaden and are reduced in height.  However, even above
$T_{c}$, the LDOS is still suppressed at low energies, in comparison
to the normal state.  This result agrees with a previous study
\cite{eckl, eckl2} which considered thermal fluctuations of the phase but not
of the magnitude of the complex pairing-field, and included no
quenched disorder.

We have also studied the $T$-dependence of the magnitude of the pairing
field, its thermal fluctuations, and the effective superfluid
density of our disordered system.  We find that the phase ordering
temperature is greatly  reduced from the spatial average of the mean-field transition
temperatures appearing in the Ginzburg Landau free energy functional. This
reduction is due to both thermal fluctuations and quenched disorder in our model.

Although our work involves a non-self-consistent solution of a
$d$-wave BCS Hamiltonian, it differs from previous studies of this
kind\cite{fang, ghosal, ghosal1, cheng_su} because it includes
thermal fluctuations as well as  quenched disorder in the
pairing-field amplitude. For our model, quenched disorder is crucial in
obtaining LDOS spectra which depend smoothly on energy and are also
consistent with the observed low and broad peaks in the $\beta$
regions.

The rest of this article is organized as follows:  In Section II, we
present the BCS model Hamiltonian.  In Section III, we derive the
discrete form of the Ginzburg Landau free energy functional used in
our calculations.  We also discuss simple estimates of the phase
ordering temperature, our choice of model parameters and our method
of introducing inhomogeneities into our model. Section IV describe
the computational methods used at both zero and finite temperature.
These methods include a classical Monte Carlo approach to treat
thermal fluctuations, exact diagonalization to obtain the LDOS, and
the reduction of finite size effects on the LDOS by the inclusion of
a magnetic field.   Section V presents our numerical results at both
$T = 0$ and finite $T$.  A concluding discussion and summary are
given in Section VI.


\section{Model}

\noindent
\subsection{Microscopic Hamiltonian}
We consider the following Hamiltonian:
\begin{equation}
  H_{BCS} =   2 \sum_{\langle i,j \rangle,\sigma}t_{ij}c_{i\sigma}^{\dagger}c_{j\sigma}
+2\sum_{\langle i,j \rangle}(\Delta_{ij}c_{i\downarrow}c_{j\uparrow} +
\text{c.c.}) -\mu \sum_{i,\sigma}c_{i\sigma}^{\dagger}c_{i\sigma}
  \label{eq:hamil_bcs}
\end{equation}
Here, $\sum_{\langle i, j \rangle }$ denotes a sum over distinct pairs
of nearest neighbors
on a square lattice with $N$ sites, $c_{j\sigma}^{\dagger}$ creates an
electron with spin $\sigma$ ($\uparrow$ or $\downarrow$) at site $j$,
$\mu$ is the chemical potential, $\Delta_{ij}$ denotes the strength of
the pairing interaction between sites $i$ and $j$, and $t_{ij}$ is the
hopping energy, which we write as
\begin{equation}
  t_{ij} = -t_{hop}.   
  \label{eq:hopp_const}
\end{equation}
where $t_{hop} > 0$.

Following a similar approach to that of Eckl {\it et
al.}~\cite{eckl}, we take $\Delta_{ij}$ to be given by
\begin{equation}
  \Delta_{ij} = \frac{1}{4}\frac{|\Delta_i|+|\Delta_j|}{2} e^{i
  \theta_{ij}},
  \label{delta}
\end{equation}
where
\begin{equation}
  \theta_{ij} =
  \begin{cases}
    (\theta_i+\theta_j)/2, & \text{if bond $\langle i, j \rangle$ is in $x$-direction,}\\
    (\theta_i+\theta_j)/2 + \pi, & \text{if bond $\langle i, j \rangle$ is in $y$-direction,}
  \end{cases}
  \label{eq:thetaij}
\end{equation}
and
\begin{equation}
  \Delta_{j}=|\Delta_j|e^{i\theta_{j}},
  \label{eq:psi_complex}
\end{equation}
is the value of the complex superconducting order parameter at site
$j$.  We will refer to the lattice over which the sums
in~(\ref{eq:hamil_bcs}) are carried out as the {\em atomic} lattice
(in order to distinguish it from the {\em XY} lattice, which will be
described in the next section.) The first term in
Eq.~(\ref{eq:hamil_bcs}) thus corresponds to the kinetic energy, the
second term is a BCS type of pairing interaction with $d$-wave
symmetry, and the third term is the energy associated with the
chemical potential.

Eq.\ (\ref{eq:hamil_bcs}) may also be written
\begin{equation}
  H_{BCS} = \Psi^{\dagger} \hat A \ \Psi - N \mu,
  \label{eq:hamil_bcs2}
\end{equation}
where
\begin{equation}
  \Psi \equiv
  \left(
    \begin{array}{c}
      c_{i\uparrow} \\
      c_{i\downarrow}^{\dagger}
    \end{array}
  \right), \quad \text{$i=1,N$}
  \label{eq:Psi}
\end{equation}
and
\begin{equation}
 \hat A= \left[
    \begin{array}{cc}
      \hat t & \hat\Delta^{\ast} \\
      \hat\Delta & -\hat t^{\ast}
    \end{array}
  \right].
  \label{eq:A}
\end{equation}
Here $\hat t$ and $\hat\Delta$ are $N \times N$ matrices with
elements $\hat t_{ij}$ [$\hat t_{ij}=t_{ij}$, as given by eq.\
(\ref{eq:hopp_const}) if $i$ and $j$ are nearest-neighbors, $\hat
t_{ij}=-\mu$ if $i=j$, and $\hat t_{ij}=0$ otherwise] and $\hat
\Delta_{ij}$ [$\hat \Delta_{ij} = \Delta_{ij}$, as given by eq.\
(\ref{delta}) if $i$ and $j$ are nearest-neighbors, and $\hat
\Delta_{ij} = 0 $ otherwise].

Let $\hat U$ be the unitary matrix that diagonalizes $\hat A$, i.\ e.,
\begin{equation}
 \hat B = \hat U^{\dagger} \hat A\ \hat U, \quad\quad \hat B \quad \text{diagonal.}
  \label{eq:B}
\end{equation}
We can then rewrite (\ref{eq:hamil_bcs2}) as
\begin{equation}
  H_{BCS} = \Phi^{\dagger} \hat B \ \Phi - N \mu,
  \label{eq:hamil_bcs3}
\end{equation}
with $\Phi$ defined by
\begin{equation}
  \Psi = \hat U\ \Phi.
  \label{eq:phi}
\end{equation}
If we make the following definitions:
\begin{equation}
  \Phi \equiv
  \left(
    \begin{array}{c}
      \gamma_{i\uparrow} \\
      \gamma_{i\downarrow}^{\dagger}
    \end{array}
  \right), \quad \text{$i=1,N$}
  \label{eq:phi2}
\end{equation}
and
\begin{equation}
 \hat U \equiv \left[
    \begin{array}{cc}
      u_{j}(r_{i})  & -v_{j}^{\ast}(r_{i}) \\
      v_{j}(r_{i}) & u_{j}^{\ast}(r_{i})
    \end{array}
  \right], \quad \text{$i$, $j=1,N$,}
  \label{eq:U}
\end{equation}
where $i$ labels the {\em row} and $j$ the {\em column} of $N \times
N$ matrices, then we can see that (\ref{eq:phi}) is the typical
Bogoliubov - de Gennes transformation~\cite{tinkham,degennes}:
\begin{eqnarray}
  c_{i\uparrow} = \sum_{j=1}^{N}[ \gamma_{j\uparrow}u_{j}(r_{i}) - \gamma_{j\downarrow}^{\dagger}v_{j}^{\ast}(r_{i})],\nonumber \\
  c_{i\downarrow} = \sum_{j=1}^{N}[ \gamma_{j\downarrow}u_{j}(r_{i}) + \gamma_{j\uparrow}^{\dagger}v_{j}^{\ast}(r_{i})].
  \label{eq:degennes}
\end{eqnarray}
Thus $\Phi$ is a $2N$-dimensional column matrix and $\hat U$ is a $2N
\times 2N$-dimensional square matrix.

Denoting the diagonal elements of the matrix $\hat B$ by $E_n$, we
can use (\ref{eq:A}), (\ref{eq:B}) and (\ref{eq:U}) to obtain
\begin{equation}
 \left[
    \begin{array}{cc}
      \hat t & \hat\Delta^{\ast} \\
      \hat\Delta & -\hat t^{\ast}
    \end{array}
  \right]
  \left[
    \begin{array}{c}
      u_{n}(r_{i}) \\
      v_{n}(r_{i})
    \end{array}
  \right]
  = E_{n}
  \left[
    \begin{array}{c}
      u_{n}(r_{i}) \\
      v_{n}(r_{i})
    \end{array}
  \right].
  \label{eq:matrix_eigenvalue}
\end{equation}
Eq.~(\ref{eq:matrix_eigenvalue}) is the eigenvalue problem which
must be solved in order to compute the local density of states, as
we describe next.

\noindent
\subsection{Explicit expression for the local density of states}

We wish to compute the local density of states, denoted LDOS$(\omega,
r_i)$, as a function of the energy $\omega$ and lattice position $r_i
= (x_i, y_i)$ at both zero and finite temperature $T$.
Given the value of the of the superconducting order parameter
$\Delta_{i}$ at each lattice site, the matrix $\hat \Delta$ can be
constructed and the $\text{LDOS}(\omega,r,\{\Delta_{i}\})$ can be
computed through~\cite{ghosal}
%
%
\begin{eqnarray}
  \text {LDOS} (\omega,r_{i},\{ \Delta_{i} \} ) = \sum_{n, E_{n} \geq 0} [
   |u_{n}(r_{i})|^2 \delta(\omega-E_{n}) + |v_{n}(r_{i})|^2 \delta(\omega+E_{n}) ]
%
  \label{eq:finite_temp_dos}
\end{eqnarray}
%
%
%
At $T=0$ all the phases $\theta_{i}$ are the same, since this choice
minimizes the energy of the superconducting system.
Thus, in this case, once we know $\{|\Delta_{i}|\}$ we can solve
eq.~(\ref{eq:matrix_eigenvalue}) for $u_{n}(r_{i})$, $v_{n}(r_{i})$
and $E_{n}$, and use this solution in (\ref{eq:finite_temp_dos}). At
finite $T$, since $\Delta_{i}$ will thermally fluctuate, we need a
procedure to obtain an average of (\ref{eq:finite_temp_dos}) over the
relevant configurations of $\{\Delta_{i}\}$. We explain that procedure
next.

\section{Model for thermal fluctuations}

At finite $T$ we compute LDOS$(\omega,r_i)$ by performing an average
of $\text {LDOS} (\omega,r_{i},\{\Delta_{i}\})$ over different
configurations $\{\Delta_{i}\}$. Those configurations are obtained
assuming that the thermal fluctuations of $\{\Delta_{i}\}$ are
governed by a Ginzburg-Landau (GL) free energy functional $F$, which
is treated as an effective classical Hamiltonian.
%
%
%

The Ginzburg Landau free energy functional has been widely studied and
applied to a variety of systems. It has been extensively used to study
granular conventional superconductors
\cite{muhlschlegel_scalapino_denton, fisher_barber_jasnow,
  deutscher_imry_gunter, patton_lamb_stroud, ebner_stroud23,
  ebner_stroud25, ebner_stroud28, ebner_stroud39}.  Other studies have
focused on the use of GL theory to describe the phase diagram of
extreme type II superconductors \cite{nguyen_sudbo}, the influence
of defects on the structure of the order parameter of $d$-wave
superconductors \cite{xu_ren_ting, alvarez_buscaglia_balseiro}, and
the effect of thermal fluctuations
on the heat capacity of high temperature superconductors
\cite{ebner_stroud39, ramallo_vidal}.  Yet other researchers have
derived the GL equations for vortex structures from microscopic
theories \cite{ren_xu_ting}. There has also been interest studying the
nature of the transition in certain parameter ranges for this type of
model \cite{alvarez_fort, bittner_janke}.

In this section we discuss a procedure for obtaining a suitably
discrete form of $F$, and determining its coefficients from
experiments.  [The final form of $F$ is given by eq.\
(\ref{eq:GLgeneral4}).]  We also discuss a way to estimate the phase
ordering temperature using this model, the choice of the parameters
that determine the GL coefficients, and finally a method of introducing
inhomogeneities into the model.


\noindent
\subsection{Discrete form of the Ginzburg-Landau free energy}

For a continuous superconductor in the absence of a vector
potential, the Ginzburg-Landau free energy density has the
form
\begin{equation}
  F^{\prime}= \alpha \left(\frac{T}{T_{c0}}-1\right)|\psi^{\prime}|^2 +
  \frac{b}{2} |\psi^{\prime}|^4 + \frac{\hbar}{2m^{\ast}}|\nabla\psi^{\prime}|^2.
  \label{eq:energy_density}
\end{equation}
Since $|\psi^{\prime}|^2$ and F$^\prime$ have dimensions of inverse volume,
and energy per unit volume, it
follows that $\alpha$ and $b$ have dimensions of energy, and (energy
$\times$ volume), respectively.

The squared penetration depth $\lambda^2(T)$ and zero-temperature
Ginzburg-Landau coherence length $\xi_{0}$ are related to the
coefficients of $F$ by~\cite{tinkham}
\begin{equation}
  \alpha = \frac{\hbar^2}{2m^{\ast}\xi_{0}^{2}},
  \label{eq:alpha}
\end{equation}
and
\begin{equation}
  b = 8 \pi \mu_{B}^2\left(\frac{\lambda(0)}{\xi_{0}}\right)^2
  \label{eq:b}
\end{equation}
where $\mu_{B}^2\simeq 5.4\times 10^{-5}\text{eV}$-$\AA ^3$ is the
square of the Bohr magneton.

Let us assume that the position-dependent superconducting energy gap
$\Delta_{i}$ at $r_i$ is related to $\psi^{\prime}_{i}$, as in
conventional BCS theory,
 through
\begin{equation}
|\psi^{\prime}_{i}|^2 =
\frac{\alpha_{i}}{9.38b_{i}}\left|\frac{\Delta_{i}}{k_{B}T_{c0i}}\right|^2,
  \label{eq:psi_delta}
\end{equation}
where we have also assumed that $T_{c0}$, $b$, and $\alpha$ are
functions of position.   The validity of (\ref{eq:psi_delta}) can be
verified by noting that in the absence of fluctuations
$F^{\prime}$ is minimized by
\begin{equation}
|\psi^{\prime}_{i}|^2 =
\frac{\alpha_{i}}{b_{i}}\left(1-\frac{T}{T_{c0i}}\right).
  \label{eq:minimized}
\end{equation}
Combining (\ref{eq:minimized}) with (\ref{eq:psi_delta}), we obtain,
at $T=0$,
\begin{equation}
|\Delta_{i}(0)|^2 = 9.38 (k_{B}T_{c0i})^2.
  \label{eq:zero_temp_gap}
\end{equation}
This result agrees well with experiment provided (i)  $T_{c0}$ is
interpreted as the temperature at which an energy gap opens
according to ARPES experiments, and (ii) $\Delta(0)$ is
taken as the low-temperature ($T<<T_c$)  magnitude of the gap
observed in ARPES and tunneling experiments \cite{mourachkine}.

In order to obtain a discrete version of the free energy functional,
we integrate the free energy density (\ref{eq:energy_density}) over
volume to yield the free energy
\begin{equation}
  F=\int F^{\prime}dV.
\label{eq:integrate_Fprime}
\end{equation}
Assuming that $\psi^{\prime}\sim \text{constant}$ within
a volume $\xi_{0}^{2}d$ (where $\xi_{0}$ is the zero-temperature
coherence length and $d$ is thickness of the superconducting layer),
we can discretize the layer into $M$ cells of volume $\xi_{0}^{2}d$.
Using (\ref{eq:psi_delta}), we can then write
\begin{eqnarray}
  \frac{F}{K_{1}} =
  \sum_{i=1}^{M} \left(\frac{T}{T_{c0i}}-1\right)\frac{1}{\lambda^{2}_{i}(0)} \left|\frac{\Delta_{i}}{k_{B}T_{c0i}}\right|^2 +
  \sum_{i=1}^{M} \frac{1}{2(9.38)}\frac{1}{\lambda^{2}_{i}(0)} \left|\frac{\Delta_{i}}{k_{B}T_{c0i}}\right|^4 \nonumber \\
  + \sum_{\langle ij \rangle}\left| \frac{\Delta_{i}}{\lambda_{i}(0)k_{B}T_{c0i}} -
\frac{\Delta_{j}}{\lambda_{j}(0)k_{B}T_{c0j}} \right|^2,
\label{eq:general1}
\end{eqnarray}
where
\begin{equation}
  K_{1} \equiv \frac{\hbar^4 d}{32 (9.38) \pi m^{\ast 2} \mu_{B}^{2}}
\end{equation}
`$K_{1}\simeq 2866$ eV-$\AA^2$ if $d=10\AA$. Except for $d$, $K_{1}$
is independent of material-specific parameters.

In (\ref{eq:general1}) the sums are performed over what we will call
the {\em XY} lattice, which is not necessarily the same as the {\em
  atomic} lattice used in (\ref{eq:hamil_bcs}).
In~(\ref{eq:general1}), $\Delta_{i} = |\Delta_{i}|e^{-i\theta_{i}}$ is
the value of the superconducting order parameter on the $i$th cell of
the {\em XY} lattice. The third sum is carried out over {\em distinct}
pairs of nearest-neighbors cells $\langle ij \rangle$.

In order to see how the {\em XY} lattice and the atomic lattice are
related, we now analyze some of the relevant length scales in our
problem. Typically, the linear dimension of the {\em XY} lattice
cell is taken to be the $T = 0$ coherence length $\xi_{0}$ of the
material in the superconducting layer.  In a cuprate superconductor,
e. g.,
Bi2212, $\xi_{0} \approx 15 \AA$, while the lattice constant of
the microscopic (atomic) Hamiltonian of eq.\ (1) - i.\ e., the
distance between the Cu sites in the CuO$_2$ plane - is $a_{0}
\approx 5.4 \AA$~\cite{mourachkine}.
Thus, in this case, a single {\em XY} cell would contain about nine
sites of the atomic lattice, on each of which the superconducting
order parameter would have the same value $\Delta_i$.

It is convenient to introduce a dimensionless superconducting gap
\begin{equation}
  \psi_{i} \equiv \frac{\Delta_{i}}{E_{0}},
  \label{eq:psi_def}
\end{equation}
and a dimensionless temperature
\begin{equation}
  t \equiv \frac{k_{B}T}{E_{0}},
  \label{eq:t_def}
\end{equation}
where $E_{0}$ is an arbitrary energy scale which will be specified
below.  We can then rewrite (\ref{eq:general1}) as
\begin{eqnarray}
  \frac{F}{K_{1}} =
  \sum_{i=1}^{M} \left(\frac{t}{t_{c0i}}-1\right)\frac{1}{\lambda^{2}_{i}(0)t_{c0i}^2} \left|\psi_{i}\right|^2 +
  \sum_{i=1}^{M} \frac{1}{2(9.38)}\frac{1}{\lambda^{2}_{i}(0)t_{c0i}^4} \left|\psi_{i}\right|^4  \nonumber \\
  + \sum_{\langle ij \rangle}
  \left[
      \left| \frac{\psi_{i}}{\lambda_{i}(0)t_{c0i}} \right|^2 +
      \left| \frac{\psi_{j}}{\lambda_{j}(0)t_{c0j}} \right|^2 -
      \frac{2|\psi_{i}||\psi_{j}|}{\lambda_{i}(0)t_{c0i}\lambda_{j}(0)t_{c0j}}
      \cos(\theta_{i}-\theta_{j}).
    \right]
\label{eq:GLgeneral3}
\end{eqnarray}
In our calculations, we will employ periodic boundary conditions. In
that case, sums of the form $\sum_{\langle ij \rangle}(a_i+a_j)$ can
be replaced by $4\sum_{i} a_i$, and
\begin{eqnarray}
    \frac{F}{K_{1}} =
    \sum_{i=1}^{M} \left(\frac{t}{t_{c0i}}+3\right)\frac{1}{\lambda^{2}_{i}(0)t_{c0i}^2} \left|\psi_{i}\right|^2 +
    \sum_{i=1}^{M} \frac{1}{2(9.38)}\frac{1}{\lambda^{2}_{i}(0)t_{c0i}^4} \left|\psi_{i}\right|^4  \nonumber \\
    -\sum_{\langle ij \rangle}
    \frac{2|\psi_{i}||\psi_{j}|}{\lambda_{i}(0)t_{c0i}\lambda_{j}(0)t_{c0j}}
    \cos(\theta_{i}-\theta_{j}).
  \label{eq:GLgeneral4}
\end{eqnarray}
Eq.\ (\ref{eq:GLgeneral4}) is the most general form for the
Ginzburg-Landau free energy functional considered in our
calculations.
In our simulations we allow both the amplitude $|\psi |$ and the phase
$\theta$ of $\psi$ to undergo thermal fluctuations.


\subsection{Thermal averages}
As mentioned at the beginning of this section, at finite $T$ we
compute LDOS$(\omega,r_i)$ by performing an average of $\text {LDOS}
(\omega,r_{i},\{\psi_{i}\})$ over different configurations
$\{\psi_{i}\}$. Those configurations are obtained assuming that the
thermal fluctuations of $\{\psi_{i}\}$ are governed by the
Ginzburg-Landau (GL) free energy functional $F$ described above.  $F$
is treated as an effective classical Hamiltonian, and thermal averages
$\langle...  \rangle$ of quantities $Q$, such as LDOS$(\omega,r_i)$,
are obtained through
\begin{equation}
  \langle Q \rangle = \frac{\int \prod_{i=1}^{N} d^2\psi_{i}\, e^{ -F/ k_{B}T} Q(\{\psi_i \}) }{Z},
  \label{eq:Q_thermal_avg}
\end{equation}
where $Z$ is the canonical partition function,
\begin{equation}
  Z = \int \prod_{i=1}^{N} d^2\psi_{i}\, e^{ -F /k_{B}T}.
  \label{eq:partition_function}
\end{equation}

\noindent
\subsection{Estimate of the Kosterlitz-Thouless transition}

If amplitude fluctuations are neglected, the Hamiltonian
(\ref{eq:GLgeneral4}) would correspond to an XY model on a square
lattice.  If the system is homogeneous, this XY model undergoes a
Kosterlitz-Thouless transition at a temperature
\begin{equation}
  k_{B}T_{c}\simeq 0.89 J_{XY},
  \label{eq:KT_transi}
\end{equation}
where $J_{XY}$ is the coupling constant between spins:
\begin{equation}
  H_{XY} = - J_{XY} \sum_{\langle ij \rangle}
  \cos(\theta_{i}-\theta_{j}).
  \label{eq:xy}
\end{equation}
From eqs.\ (\ref{eq:GLgeneral4}) and (\ref{eq:xy}), the XY coupling
between sites $i$ and $j$ is given by
\begin{equation}
  J_{XY,ij}(t) \equiv \frac{2 K_{1} |\psi_{i}||\psi_{j}|}{\lambda_{i}(0)t_{c0i}\lambda_{j}(0)t_{c0j}}.
  \label{eq:JXY_deff}
\end{equation}
If we approximate $\psi_{i}(t)$ by the value that minimizes
$F^{\prime}$ when fluctuations are neglected,
\begin{equation}
  |\psi_{i}(t)|\simeq \sqrt{9.38 (1-t/t_{c0i})}\, t_{c0i},
  \label{eq:psi}
\end{equation}
then
\begin{equation}
  J_{XY,ij}(t) \simeq \frac{2(9.38)\sqrt{ (1-t/t_{c0i})  (1-t/t_{c0j}) } }{\lambda_{i}(0)\,\lambda_{j}(0)}.
  \label{eq:JXY1}
\end{equation}
which in the homogeneous case reduces to
\begin{equation}
  J_{XY}(t) \simeq \frac{18.76(1-t/t_{c0}) }{\lambda^{2}(0)}.
  \label{eq:JXY2}
\end{equation}
This result and eq.~(\ref{eq:KT_transi}) give
\begin{equation}
  T_{c} \simeq \frac{T_{c0}}{1 + T_{c0}/\gamma_{1}},
  \label{eq:Tc}
\end{equation}
where
\begin{equation}
  \gamma_1 = \frac{(0.89)(18.76)K_{1}}{\lambda^{2}(0)k_{B}}.
\end{equation}
Eq.\ (\ref{eq:Tc}) can also be rewritten as
\begin{equation}
  t_{c} \simeq \frac{t_{c0}}{1 + t_{c0}\gamma_2},
  \label{eq:tc}
\end{equation}
where $\gamma_2 = \frac{E_{0}}{k_{B}\gamma_1}$. Using $\lambda(0)$ =
1800 $\AA$ and $d$ = 10 $\AA$, we obtain $\gamma_1=172K$. Finally, if
we choose $E_{0}=200$ meV (for reasons given below), we obtain
$\gamma_2 = 13.54$.

Expressions (\ref{eq:Tc}) and (\ref{eq:tc}) will typically
overestimate the phase-ordering (or Kosterlitz-Thouless) transition
temperature $T_c$.  Both thermal fluctuations of $|\psi|$ and
quenched disorder will generally reduce $T_{c}$ below these
estimates.

\noindent
\subsection{Choice of parameters}

Next, we describe our choice of parameters entering both the
microscopic model [Eq.~(\ref{eq:hamil_bcs})], and that for thermal
fluctuations [Eq.~(\ref{eq:GLgeneral4})].   In a typical cuprate,
such as underdoped Bi2212, the low-$T$ superconducting gap is $\sim
50$ meV, the hopping integral $t_{hop} \sim 200$ meV, $\lambda(0)
\sim 1800\AA$, and the pseudogap opens at $T_{c0} \sim 200K \simeq
20meV/k_{B}$.   Also, the lattice constant of the CuO$_2$ lattice
plane is $a_{0}\sim 5.4 \AA$,
 while $\xi_{0} \sim 15 \AA$. If in eqs.\
(\ref{eq:psi_def}) and (\ref{eq:t_def}) we choose $E_{0} = t_{hop} =
200$meV, then, using those expressions, we obtain $|\psi(0)| = 0.25$
and $t_{c0} = 0.1$.  We can substitute these values into eq.\
(\ref{eq:Tc}) to obtain an estimate for the phase ordering
temperature, namely $T_{c}=130K$.   Our actual simulations, carried
out in the presence of thermal fluctuations of the gap magnitude and
quenched disorder, actually yield a lower $T_{c}$, as expected.

We have carried out calculations using this set of parameters, but
also with smaller values of $\xi_0$, in order to treat larger XY
lattices.  Suppose we wish to carry out a simulation on a $16\times
16$ XY lattice. If we use the parameters values described above, we
would have a $48 \times 48$ atomic lattice.  To compute the density
of states on this lattice, we would have to diagonalize $4608 \times
4608$ matrices [see Eq.~(\ref{eq:A})].  Each such diagonalization
takes $\sim$ 1 hour on a node for serial jobs of the OSC Pentium 4
Cluster, which has a 2.4 GHz Intel P4 Xeon processor.
Since thermal averages require several hundred diagonalizations,
a $16\times 16$ XY lattice is too large using these parameters. If,
however, we choose a smaller coherence length, we will have fewer
atomic sites per XY cell, and hence a smaller  matrix to diagonalize
for a $16 \times 16$ XY lattice.
In the BCS formalism, $\xi_{0}\propto v_{F}/|\Delta|$, where $v_{F}$
is the Fermi velocity.  Thus, if $\xi_{0}$ is $n$ times smaller than
the experimental value, then, for fixed $v_F$, $\Delta$, and hence
$t_{c0}$,  will be $n$ times larger than that value.


\noindent
\subsection{Inhomogeneities}

As noted above, experiments
show that in some
cuprates the energy gap is spatially inhomogeneous \cite{cren,
  pan_davis, howald1, lang_davis, howald2, kato_sakata, fang, mashima}.
Typically, in some spatial regions, which we call $\alpha$-regions,
the LDOS has a small gap and large coherence peaks, while in other
regions, the $\beta$-regions, the LDOS has a larger gap and reduced
coherence peaks. The percentage of the area occupied by $\alpha$ and
$\beta$ regions, respectively, depends on the doping concentration. In
Bi2212, for example, a nearly optimally doped sample (hole dopant
level $\sim 0.18$) has $\sim 10$ \% of the area occupied by $\beta$
regions, while for an underdoped sample (hole dopant level $\sim
0.14$), the areal fraction of the $\beta$ regions is about $\sim 50$
\% \cite{lang_davis}.

We introduce spatial inhomogeneities into our model by including a
binary distribution of $t_{c0i}$'s.
%
%
Typically, we chose the smaller value of $t_{c0i}$ so that, for a
homogeneous system, the gap $\Delta_{i}(0)$ resulting from our model
[eq. \ref{eq:psi}] approximately equals that observed in experiments
(for further details, see discussion in the subsection entitled
``choice of parameters'').
We refer to XY cells with this small $t_{c0i}$ as $\alpha$-cells.  For
the $\beta$-cells, on the other hand, we assume a value $t_{c0i}$ $K$
times large than that of the $\alpha$-cells. We obtained our best
results by choosing $K = 3$.
%
%
We have carried out simulations considering both an ordered and a
random distribution of $\beta$-cells.
%

To determine the distribution of $\lambda_i(0)$, we use the connection
between the local superfluid density $n_{s,i}(T)$ and $\lambda_i(T)$
implied by eqs.\ (\ref{eq:alpha}), (\ref{eq:b}) and
(\ref{eq:psi_delta}):
\begin{equation}
  n_{s,i}(T) = |\psi^{\prime}_{i}(T)|^2 = \frac{\hbar^2}{(9.38)16\pi\mu_B^2 m^\ast}
  \frac{|\Delta_i(T)|^2}{(k_BT_{c0i})^2}\frac{1}{\lambda_i^2(0)}.
\end{equation}
Thus, at fixed but very low $T$, since $|\Delta_i(0)/k_BT_{c0i}|^2$
is independent of position according to our model [see
Eq.~(\ref{eq:zero_temp_gap})],
%
$n_{s,i}(T)\propto 1/\lambda_i^2(0)$. Since the coherence peaks in
the local density of states are observed to be lower where the gap
is large, we will assume that $t_{c0i}$ and $\lambda_i^2(0)$ are
correlated according to the equation
\begin{equation}
  \lambda_i^2(0) = \frac{  \lambda^2(0)  }{t_{c0}}t_{c0i}
  \label{eq:lambda2_correl}
\end{equation}
where $t_{c0}$ and $\lambda^2(0)$ are obtained from the observed
bulk properties of the material under consideration. [For example,
we typically obtain $t_{c0}$ from (\ref{eq:zero_temp_gap}) where we
take $|\Delta_{i}(0)|$ as the average of the low temperature gap
observed in experiments, and $\lambda(0)=1800\AA$.]  Substituting
(\ref{eq:lambda2_correl}) into eq.\ (\ref{eq:JXY1}) gives, for
$t<<t_{c0i}$ and  $t<<t_{c0j}$,
\begin{equation}
  J_{XY,ij} \propto \frac{1}{\sqrt{t_{c0i}\,t_{c0j}}}.
  \label{eq:JXY3}
\end{equation}
%

\section{Computational method}

\noindent
\subsection{Monte Carlo}

We compute thermal averages of several quantities, including $\text
{LDOS} (\omega,r_{i}, T)$, using a Monte Carlo (MC) technique.  Thus,
we estimate integrals of the form (\ref{eq:Q_thermal_avg}) using
\begin{equation}
  \langle  Q \rangle = \frac{1}{N_{m}} \sum_{j=1}^{N_{m}}  Q( \{ \psi_{i} \} ),
  \label{eq:Q_thermal_avg_estimator}
\end{equation}
where $N_{m}$ is the number of configurations $\{ \psi_{i} \}$ used to
compute the average, and the configurations $\{ \psi_{i} \}$ are
obtained using
the standard Metropolis algorithm \cite{barkema_newman,thijssen} as we now describe.
We first set the values of the $t_{c0i}$ and $\lambda_{i}(0)$ in
each XY lattice cell as described in the previous section. This
completely determines the GL free energy functional $F$ [Eq.\
(\ref{eq:GLgeneral4}).] We then set the initial values of $\psi_{i}$
so as to minimize $F$.  Next we perform attempts to change the value
of each $\psi_{i}$ by $\delta_i$, where $\delta_{i}$ is the complex
number $\delta_i= \delta_{i, re} + i \delta_{i,im}$, and $\delta_{i,
  re}$ and $i \delta_{i,im}$ are random numbers with a uniform
distribution in the range $[-\delta_{0},\delta_{0}]$. We define a
\emph{MC step} as an attempt to change the value $\psi_{i}$ on each of
the XY cells.  The value of $\delta_{0}$ is in turn adjusted at each
temperature so that attempts to change $\psi$ have a success rate of
50\%.
Attempts to change $\psi_{i}$ are accepted with a probability
$\exp(-\Delta F/k_{B}T)$, where $\Delta F =
F[\psi_{1},\psi_{2},\ldots,\psi_{i}+ \delta_{i},\ldots,\psi_{M}] -
F[\psi_{1},\psi_{2},\ldots,\psi_{i},\ldots,\psi_{M}]$.  In this way,
different configurations $\{ \psi_{i} \}$ are obtained.

In order to select which of those configurations $\{ \psi_{i} \}$ to
use in (\ref{eq:Q_thermal_avg_estimator}), we first made an estimate
the phase autocorrelation time $\tau$ \cite{eckl2}, in units of MC
steps, at
each temperature. We chose $\tau=\text{min}[\tau^{\prime},500]$, where
$\tau^{\prime}$ is implicitly defined by
\begin{equation}
  \frac{c(\tau^{\prime})}{c(0)} = \frac{1}{e},
  \label{eq:autocorrel1}
\end{equation}
and $c(\tau^{\prime})$ is an space average of the phase autocorrelation
function \cite{thijssen}:
\begin{equation}
  c(\tau^{\prime}) =  \frac{1}{M} \sum_{j=1}^{M}
  \left[
    \langle e^{i\theta_{j}(\tau^{\prime})} e^{-i\theta_{j}(0)} \rangle -
    \langle e^{i\theta_{j}(\tau^{\prime})} \rangle \langle  e^{-i\theta_{j}(0)} \rangle
  \right].
  \label{eq:autocorrel2}
\end{equation}
%
%
Once we estimated $\tau$, we performed 20$\tau$ MC steps to allow the
system to equilibrate, then we carried out an additional 100$\tau$ MC
steps at each $T$ for each disorder realization.
During those
100$\tau$ MC steps, we sampled $\{ \psi_{i} \}$ every $\tau$ MC step,
thus obtaining $N_{m}=100$ configurations to use in
(\ref{eq:Q_thermal_avg_estimator}) to estimate the quantities of
interest.
We also performed longer simulations, averaging over $N_{m}=300$
configurations to compute the LDOS, and $N_{m}=5000$ configurations
to compute $\gamma$, $|\psi|$ and the root-mean-square fluctuations
$[\sigma_{|\psi|}]$ (defined below), obtaining virtually the same
results as with $N_{m}=100$ configurations.

When carrying out the simulation, we need a mapping between the
sites of the XY lattice and those of the atomic lattice.  To do this
mapping, we divide the atomic lattice into  regions of area $\xi_{0}
\times \xi_{0}$.   Each such region constitutes an XY cell.  All
atomic sites within such a cell are assigned the same value of the
order parameter $\psi_i$.   Clearly, the lattice constant $\xi_0$ of
the XY lattice must be an integer multiple of the atomic lattice
constant $a_0$.
Thus, if our XY lattice has $L^2 $ sites, then the atomic lattice
has $[L\xi_{0}/a_{0}]^2$ sites.

We diagonalize all matrices numerically using LAPACK~\cite{lapack}
subroutine ``zheev'', which can find all of the eigenvectors and
eigenvalues of a complex, hermitian matrix.  We calculate the density
of states by distributing the eigenvalues into bins of width
$\Delta\omega$.
The delta function appearing in (\ref{eq:finite_temp_dos}) is
approximated by
\begin{equation}
  \delta(x) = \frac{1}{\pi}\frac{\epsilon}{\epsilon^2 + x^2},
\end{equation}
%
where we choose $\epsilon \sim \Delta\omega \sim 0.01 t_{hop}$.

\noindent
\subsection{Reducing finite size effects through inclusion of a magnetic field}

To reduce finite size effects on LDOS$(\omega,r)$, we use a method
introduced by Assaad~\cite{assaad}.  The basic idea of this method
is to break the translational invariance of $t_{hop}$ through the
substitution $t_{hop}\rightarrow t_{ij}(L)$ in
Eq.~(\ref{eq:hopp_const}).  This is done so as to improve convergence
of the quantities of interest, such as LDOS$(\omega,r)$, as a function
of the size of the atomic lattice $N$ \cite{eckl2}.
However, $t_{ij}(N)$ must still satisfy
\begin{equation}
  \lim_{N\rightarrow \infty} t_{ij}(N) = -t_{hop},
  \label{eq:hopp_const2}
\end{equation}
so that the original form of $t_{ij}$ is recovered in the
thermodynamic limit.

Assaad showed that if one makes the substitution $t_{hop}\rightarrow
t_{ij}(N)$ through the inclusion of a finite magnetic field, the
convergence of the density of states is greatly improved.  The
magnetic field enters through the Peierls phase factor:

\begin{equation}
  t_{ij} = -t_{hop}\, e^{i A_{\vec i \vec j}},
  \label{hopp_const2}
\end{equation}
with
\begin{equation}
  A_{\vec i \vec j} = \frac{2\pi}{\Phi_0}\int_{\vec i}^{\vec j}\vec A(\vec r)\cdot d\vec
  r.
\end{equation}
Here $\vec A(\vec r)$ is the vector potential at $\vec r$, $\Phi_0 =
hc/e$ is the flux quantum corresponding to one electronic charge
$e$, and the integral runs along the line from site $i$ to site $j$.

We use a gauge which allows periodic boundary conditions, and with
which the flux through the atomic lattice can be chosen to be any
integer multiple of $\Phi_0$~\cite{assaad, yu_stroud}.
Let $\vec i = (x\hat e_{x}a_{0}, y\hat e_{y}a_{0})$,
$\hat e_{x}$ and $\hat e_{y}$ are unit vectors in the $x$ and $y$
directions, and $x$ and $y$ are integers in the range $[0, N-1]$. Then
\begin{equation}
  A_{\vec i \vec j} =
  \begin{cases}
    \pm \frac{2\pi m}{N^2}x, & \text{if} \quad \vec j = \vec i \pm a_{0} \hat e_{y},\\
    - \frac{2\pi m}{N}y,   & \text{if} \quad \vec j = \vec i + a_{0} \hat e_{x} \quad \text{and}\quad x=N-1,\\
      \frac{2\pi m}{N}y,   & \text{if} \quad \vec j = \vec i - a_{0} \hat e_{x} \quad \text{and}\quad x=0,\\
      0, & \text{otherwise}
  \end{cases}
  \label{eq:delta1}
\end{equation}
%
where $m$ is the number of flux quanta through the atomic lattice.
We have chosen $m=1$, so that the magnetic field in our system has
the smallest non-zero value possible.
\section{Results}

\noindent
\subsection{Zero temperature}

\begin{figure}[ht]\centering
  {\includegraphics[width=10cm]{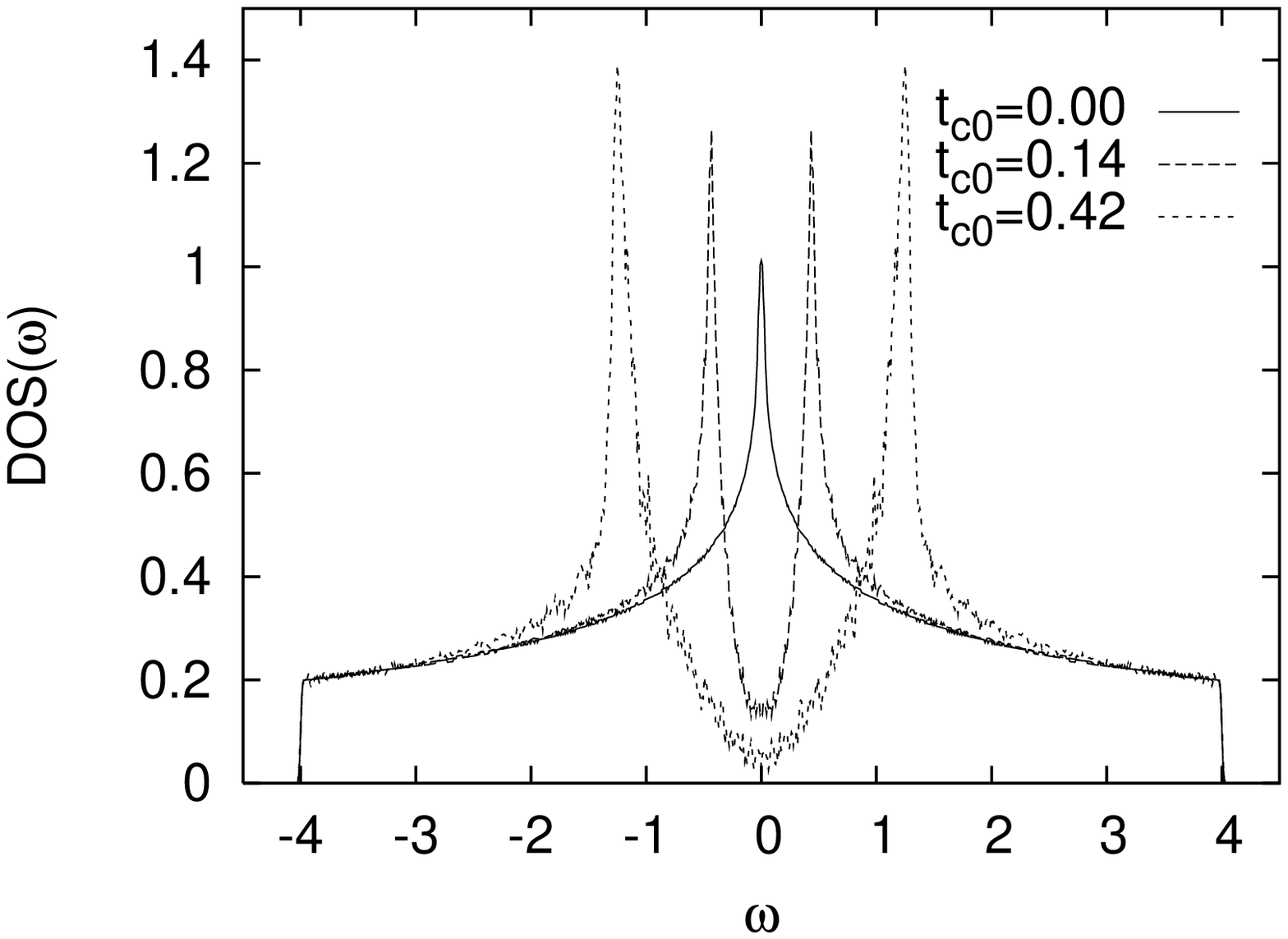}}
  \caption{
    Zero temperature density of states DOS$(\omega)$ versus energy $\omega$ for three
    homogeneous systems described by mean-field transition temperatures
    $t_{c0}=0$, $t_{c0}=0.14$, and $t_{c0}=0.42$.
    Simulations were carried out on $48 \times 48$ atomic lattices, with
    a magnetic field included, as described in the text, to reduce
    finite size effects.  Energies $\omega$ are given in units of $t_{hop}$.}
    \label{fig:ldos_pure_diff_gap_values}
\end{figure}

Fig.~\ref{fig:ldos_pure_diff_gap_values} shows the spatially
averaged density of states, DOS$(\omega)$, obtained by summing the
local density of states, LDOS$(\omega,r)$, over all sites $r$ on a
$48 \times 48$ atomic lattice with homogeneous $t_{c0}$ at
zero temperature.
%
The zero temperature
pairing strength is given by $|\psi(0)|=\sqrt{9.38}t_{c0}$, as shown
by eqs.\ (\ref{eq:zero_temp_gap}) and (\ref{eq:psi_def}).  For the
case $t_{c0}=0$, the pairing strength is zero, and we observe the
standard Van Hove peak~\cite{van_hove} for a two-dimensional
tight-binding band at $\omega = 0$.   For finite pairing strength we
observe a suppression of the density of states near $\omega = 0$,
while strong coherence peaks occur at $\omega \simeq |\psi(0)|$.


In Fig.\ 2, we compare the density of states DOS$(\omega)$ for a $32
\times 32$ atomic lattice containing a single quantum of magnetic flux
($q=1$), and a larger ($48 \times 48$) atomic lattice containing no
magnetic field ($q=0$).

 Both systems are assumed homogeneous with
$t_{c0} = 0.14$.  As can be seen, the two are very similar except at
low $|\omega|$, where the magnetic field is known to induce a change
in the density of states \cite{lages_sacramento_tesanovic}.
Note also that the zero-field DOS$(\omega)$ is less smooth than that
of the lattice with one quantum of flux, even though the zero-field
lattice is larger.  In zero-field case we have determined the
density of states using a bin width $\Delta \omega = 0.09$, while in
the finite-field case we used $\Delta \omega = 0.01$.  (The
frequencies and widths are given in units of $t_{hop}$.)   We have also carried out
a similar calculation for $q = 1$ and a $48 \times 48$ atomic lattice; the results
are similar to those shown for the $32 \times 32$ lattice except that the
density of states at $\omega = 0$ is reduced by about a third.  This Figure, and
the results just mentioned, 
show that including the magnetic field is very useful in smoothing
the density of states plots.

Before presenting our results for inhomogeneous systems, we briefly
describe our method of introducing inhomogeneities into our model.
We work with atomic lattices of size $L\times L$, in which the sites
are divided into groups of $2 \times 2$.
Each of these groups forms an XY cell, within which the
superconducting order parameter $\psi$ is kept uniform. The value of
$\psi$ in each cell is determined by the GL free energy
Eq.~(\ref{eq:GLgeneral4}), which in turn depends on the set of
values $\{t_{c0i}\}$ and $\{\lambda_{c0i}\}$.  Because $t_{c0i}$ and
$\lambda_{c0i}$ are correlated in our model, once we have the set
$\{t_{c0i}\}$, the GL free energy is completely determined and
$\psi$ at each cell can be computed through the MC method
described above.

\begin{figure}[ht]\centering
  {\includegraphics[width=10cm]{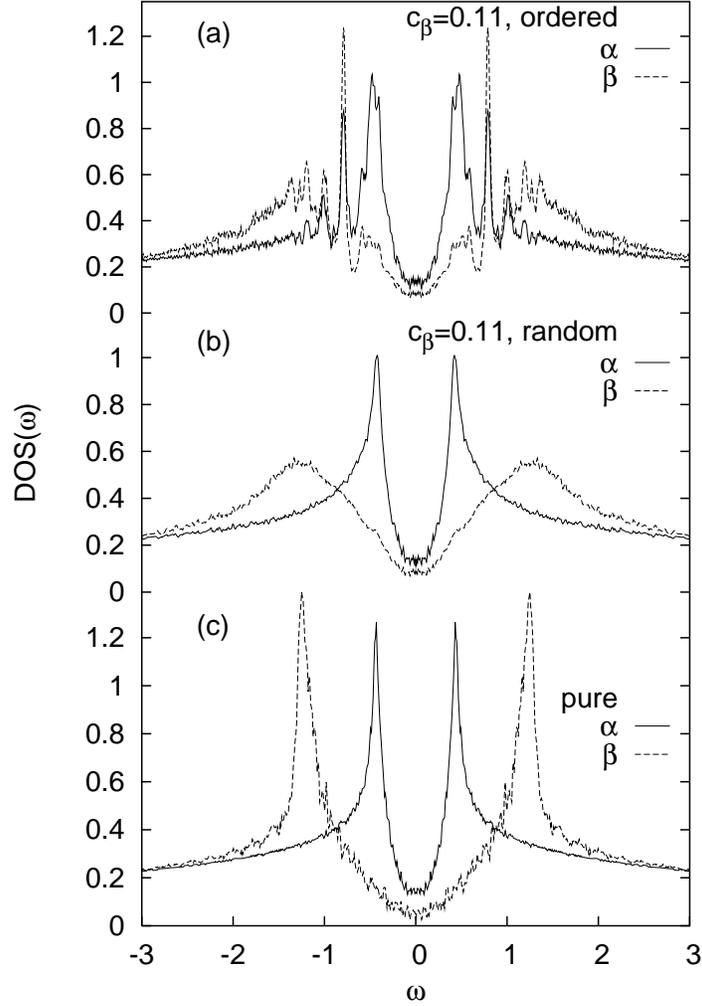}}
  \caption{
    Comparison of the zero temperature DOS of mixed $\alpha$-$\beta$
    systems (a) and (b), and pure systems (c).  In the mixed systems,
    $c_{\beta}=0.11$ is the concentration of $\beta$ cells
    ($t_{c0}=0.42$), immersed in a background of $\alpha$ cells
    ($t_{c0}=0.14$).  (a) Ordered array of $\beta$ cells; see
    Fig~\ref{fig:lattice_ord}.  (b) Disordered configuration of
    $\beta$ cells; see Fig~\ref{fig:lattice_dis}.  Curves are obtained
    by space-averaging LDOS$(\omega,r)$ over atomic
    sites within $\alpha$ or $\beta$ cells, respectively.
    In the disordered case, averages were also carried out over five different realizations
    of the disorder.  (c) DOS$(\omega)$ for two pure systems
    containing only $\alpha$ and only $\beta$ cells.
    We use a $48 \times 48$ atomic lattice;
    the size of
    an XY cell ($\alpha$ or $\beta$) is $2 \times 2$. }
    \label{fig:ldos_ord_dis_low_b}
\end{figure}

In Fig.~\ref{fig:ldos_ord_dis_low_b} (a) and (b), we show results
for two inhomogeneous systems.  Both systems consist of $48 \times
48$ atomic lattices in which a fraction $c_{\beta}=0.11$ of the XY
cells are of the $\beta$ type with $t_{c0}=0.42$, while the
remainder of the cells are of the $\alpha$ type, with $t_{c0}=0.14$.
The curves are spatial averages of the LDOS$(\omega,r)$ over the
$\alpha$ and $\beta$ cells.   In (a), they correspond to a system in
which the $\beta$ cells form an ordered array , while the curves in
part (b) correspond to a system in which the $\beta$ cells are
distributed randomly through the lattice.  For comparison,
Fig.~\ref{fig:ldos_ord_dis_low_b}(c) shows results of two
homogeneous systems: one with all $\alpha$ cells and one with all
$\beta$ cells.

The dotted line in Fig.~\ref{fig:ldos_ord_dis_low_b}(a) represents
an average of LDOS$(\omega,r)$ over the $\beta$ cells. It differs
significantly from the $\beta$ curve of the homogeneous case,
[dotted curve in Fig.~\ref{fig:ldos_ord_dis_low_b}(c)].
Specifically, instead of the single sharp, and much higher, peak in
the homogeneous $\beta$ case, there is a lower peak which is shifted
slightly to smaller $|\omega|$ and also has strong oscillations as a
function of $\omega$ (probably because of the ordered arrangement of
the $\beta$ cells).  The largest maximum of this oscillating peak is
quite sharp, however, and occurs at a distinctly smaller energy than
in the homogeneous case.

The solid line in Fig.~\ref{fig:ldos_ord_dis_low_b}(a) corresponds
to an average of the LDOS$(\omega,r)$ over $\alpha$ cells. It
differs less from the homogeneous $\alpha$ system [solid curve in
Fig.~\ref{fig:ldos_ord_dis_low_b}(c)] than in the $\beta$ case: the
main peak is not much shifted in energy, and it is slightly lower
and broader than the homogeneous case.  However, an additional peak
does appear at the same position as the larger peak of the
inhomogeneous $\beta$ curve described above.

In Fig.~\ref{fig:ldos_ord_dis_low_b}(b), we show the corresponding
density of states plots for a system with randomly distributed
$\beta$ cells. In this case we observe that the LDOS$(\omega,r)$,
averaged over $\alpha$ cells, has slightly lower and broader peaks
than that of the homogeneous $\alpha$ system shown in
Fig.~\ref{fig:ldos_ord_dis_low_b}(c), but the peaks still occur at
the same energy in both cases: $\omega \sim 0.42$.  However, the
average of the LDOS$(\omega,r)$ over the $\beta$ cells is
drastically different from the homogeneous $\beta$ case: the main
peak is greatly broadened, compared to the homogeneous $\beta$ case.

\begin{figure}[t]\centering
  {\includegraphics[width=7cm]{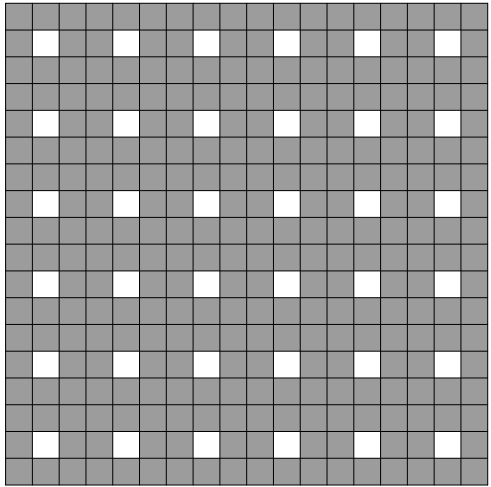}}
  \caption{
    An $18 \times 18$ XY lattice in which the $\beta$ cells form an
    ordered array: $\beta$ cells (white squares) are immersed in a
    background of $\alpha$ cells (gray squares). Each XY cell
    corresponds to an area of $\xi_{0} \times \xi_{0}$, and contains
    {\em four} atomic sites.}
  \label{fig:lattice_ord}
\end{figure}

\begin{figure}[t]\centering
  {\includegraphics[width=7cm]{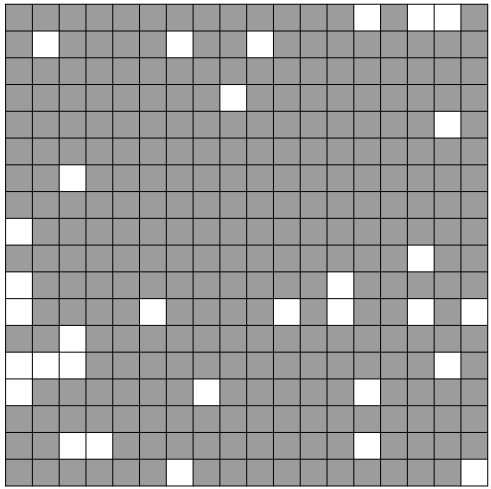}}
  \caption{
    An $18 \times 18$ XY lattice with a disordered arrangement of
    $\beta$ cells (white squares) immersed in a background of $\alpha$
    cells (gray squares). Each cell corresponds to an area of $\xi_{0}
    \times \xi_{0}$, and contains {\em four} atomic sites. This Figure
    contains a particular realization of disorder.   Density of states
    results for disordered systems are averaged over five different
    disorder
    realizations.}
  \label{fig:lattice_dis}
\end{figure}

In Figs.\  \ref{fig:lattice_ord} and \ref{fig:lattice_dis}, we show
representative ordered and disordered arrangements of $\alpha$ and
$\beta$ cells (for an $18\times 18$ XY lattice), similar to those
used in the calculations of Fig.\ 3(a) and 3(b).  In our density of
states calculations for disordered arrangements, we typically
average over about five realizations of the disorder, and use $24
\times 24$ XY lattices rather than the $18\times 18$ shown in the
schematic picture.

\begin{figure}[ht]\centering
  {\includegraphics[width=10cm]{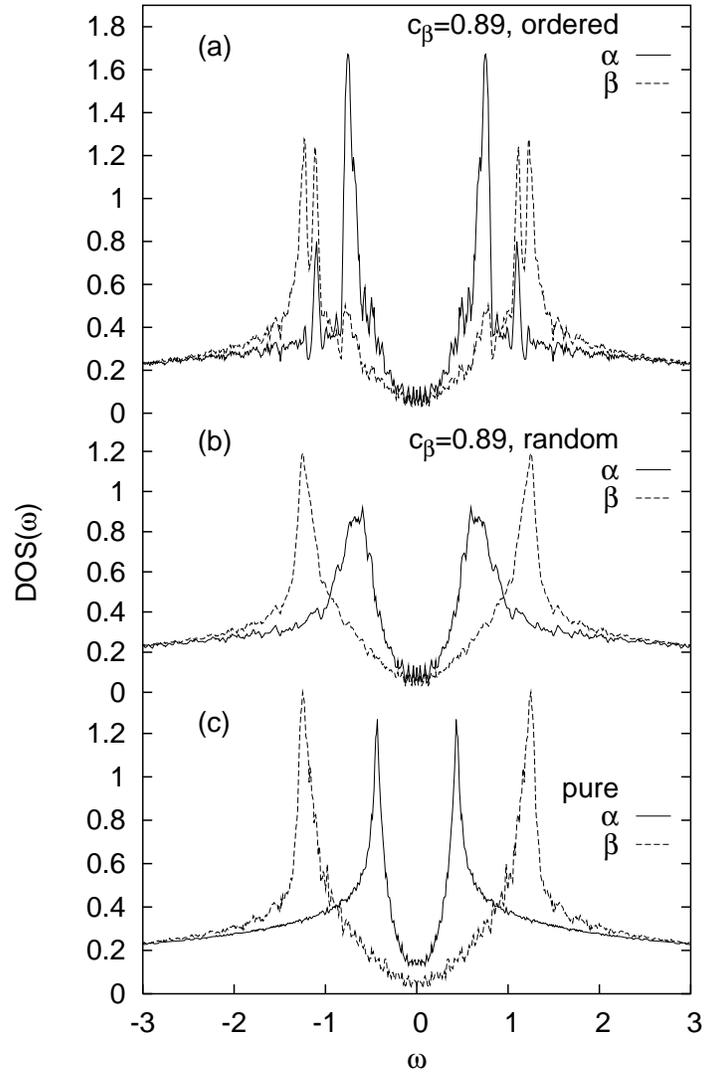}}
  \caption{
    Same as Fig.~\ref{fig:ldos_ord_dis_low_b} but with a high
    concentration $c_{\beta}=0.89$ of $\beta$ cells in the mixed
    systems.  The ordered configuration corresponds to an ordered
    arrangement of $\alpha$ cells within $\beta$.}
    \label{fig:ldos_ord_dis_low_a}
\end{figure}

In Fig.\ \ref{fig:ldos_ord_dis_low_a}, we show plots analogous to
Fig.\ \ref{fig:ldos_ord_dis_low_b}, but for a much larger
concentration of $\beta$ cells ($c_\beta = 0.89$).  Part (a) shows
results for an ordered array of $\alpha$ cells immersed in a
background of $\beta$ cells.   The simple, sharp peaks of the
homogeneous $\beta$ case [dotted curves in
Fig.~\ref{fig:ldos_ord_dis_low_b}(c) and Fig.~\ref{fig:ldos_ord_dis_low_a}(c)]
are split into two sharp peaks
at a slightly smaller energy, while the sharp peaks of the
homogeneous alpha regions, [solid curve in
Fig.~\ref{fig:ldos_ord_dis_low_b}(c)] become even sharper and
shifted toward higher energies, leading to a reduction in the
density of states near $\omega = 0$.  Also, in the inhomogeneous
$\beta$ curve of Fig.~\ref{fig:ldos_ord_dis_low_a}(a), a weak
second peak appears at the same energy as one of the peaks in the
inhomogeneous $\alpha$ curve.

The case of a disordered distribution of $\alpha$ regions immersed
in a background of $\beta$ regions is shown in
Fig.~\ref{fig:ldos_ord_dis_low_a} (b).   The peaks of the curves
corresponding to both the $\alpha$ and $\beta$ regions become lower
and broader than in the homogeneous cases,
Fig.~\ref{fig:ldos_ord_dis_low_a}(c).  The peak in the $\beta$ curve
occurs at approximately the same energy as in the homogeneous case.
The corresponding inhomogeneous $\alpha$ peak, on the other hand,
occurs at a higher energy relative to the homogeneous case.

\noindent
\subsection{Finite temperatures}

\begin{figure}[ht]\centering
  {\includegraphics[width=10cm]{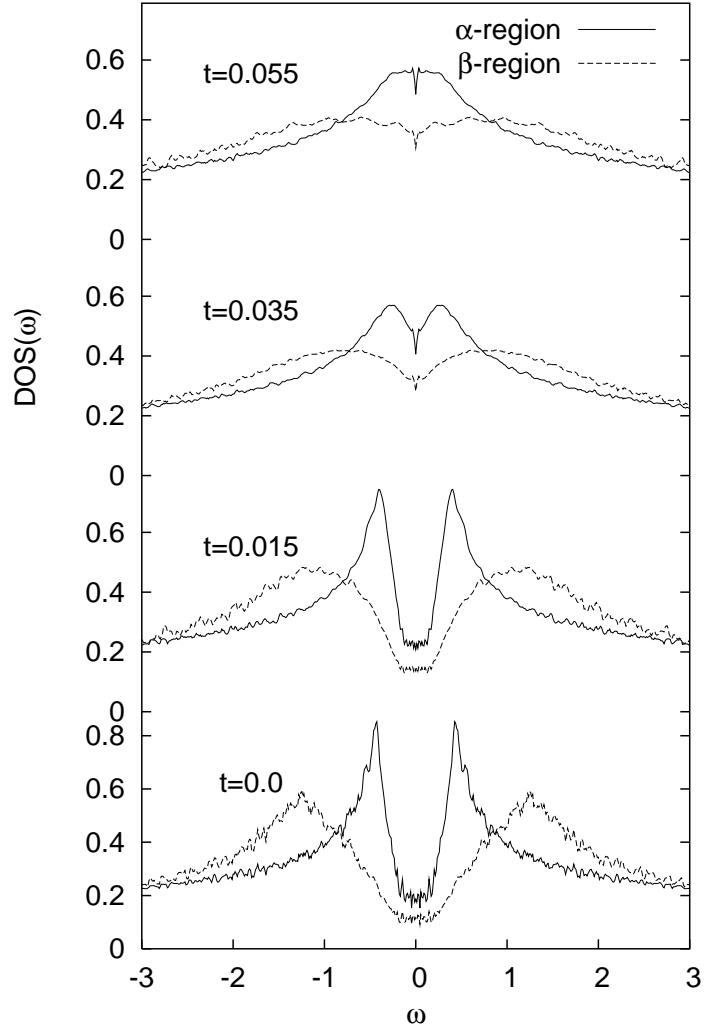}}
  \caption{
    Spatial average, at several different temperatures $t$, of the local
    density of states LDOS$(\omega,r)$ over two different types of
    cells: the $\alpha$-cells, where $t_{c0}=0.14$, and the
    $\beta$-cells, where $t_{c0}=0.42$. The $\beta$-cells occupy 10\%
    ($c_{\beta}=0.1$) of the total area, while the $\alpha$-cells
    occupy the rest. The simulations were performed using a $32\times
    32$ atomic lattice; the XY cells are $2\times 2$ atomic cells.
    The phase ordering temperature for this system is
    $t_{c}\approx 0.03$ (see the curve corresponding to
    $c_{\beta}=0.1$ in Fig.~\ref{fig:hel}).}
    \label{fig:ldos_allT}
\end{figure}

We have carried out a finite-$T$ study for the system topology most
similar to the experimental one\cite{lang_davis}: a random
distribution of $\beta$ regions immersed in a background of $\alpha$
regions.
Calculated results for such a system at $T = 0$ are shown in
Fig.~\ref{fig:ldos_ord_dis_low_b}(b). Because more matrix
diagonalizations are required at finite $T$ to obtain the relevant
thermal averages, we work with $32 \times 32$ atomic lattices,
instead of the $48 \times 48$ used at $T = 0$. Since the
computational time needed for one diagonalization scales with the
linear size $L$ of the system like $L^{6}$, each diagonalization
takes about one-tenth the time in these smaller system. Fortunately,
the reduction of finite-size effects achieved by introducing a
magnetic field leads to good results even for this relatively small
system size.   This can be seen by comparing the $t=0$ results in
Fig.~\ref{fig:ldos_allT}, which are obtained for a $32 \times 32$
atomic lattice, to the corresponding results shown
Fig.~\ref{fig:ldos_ord_dis_low_b}(b) for a $48 \times 48$ atomic
lattice.

Besides the partial densities of states, we calculate several
additional quantities at finite $t$: the effective superfluid density
$\gamma(t)$, the thermal- and space-averaged values of $|\psi|$
in the $\alpha$ and $\beta$ regions, and the relative fluctuations
$\sigma_{|\psi|}$ of $|\psi|$ averaged over each of those regions.

We compute the superfluid density $\gamma$ by averaging the diagonal
elements $\gamma_{\alpha\alpha}$ ($\alpha = x,y$) of the helicity
modulus tensor $\hat \gamma$.  Thus, we compute $\gamma =
(\gamma_{xx}+\gamma_{yy})/2$, where~\cite{ebner_stroud28}
\begin{eqnarray}
  \gamma_{xx} =
  \frac{1}{M}  \langle \sum_{\langle i,j\rangle} (x_i-x_j)^2 J_{XY,ij} \cos(\theta_{i}-\theta_{j}) \rangle -
  \frac{1}{M t} \langle  [ \sum_{\langle i,j\rangle} (x_i-x_j) J_{XY,ij} \sin(\theta_{i}-\theta_{j})]^{2} \rangle  \nonumber \\
  + \frac{1}{M t} \langle  \sum_{\langle i,j\rangle}(x_i-x_j) J_{XY,ij} \sin(\theta_{i}-\theta_{j}) \rangle^{2}.
  \label{helicity}
\end{eqnarray}
Here $x_i$ is the $x$ coordinate of $i$th XY cell $i$, $M$ is the
total number of XY cells, $J_{XY,ij}$ is the effective XY coupling
between XY cells and is given by Eq.~(\ref{eq:JXY_deff}), $\theta_i$
is the phase of $\psi_i$ and $\langle \rangle$ denotes a canonical
average.  $\gamma_{yy}$ is defined by the analogous expression with
$x_i$ replaced by $y_i$.  In our computations, we have set the
lattice constant $a_{XY}$ of the XY lattice to be unity.

The mean-square order parameter averaged over the $\alpha$ region is
computed from
\begin{equation}
  [\langle|\psi|^{2}\rangle]_{\alpha} =
  \frac{1}{M_{\alpha}} \sum_{i\in\alpha} \langle|\psi_{i}|^{2}\rangle,
  \label{eq:psi_avg}
\end{equation}
where the sum is carried out over all $M_{\alpha}$ XY cells of type
$\alpha$. $\left[\langle|\psi|^{2}\rangle\right]_{\beta}$ is defined
similarly.   The mean magnitude of the order parameter in the $\alpha$
and $\beta$ regions, denoted $[\langle|\psi|\rangle]_\alpha$ and
$[\langle|\psi|\rangle]_\beta$, are defined by an equation analogous
to eq.\ (\ref{eq:psi_avg}).
We compute the relative fluctuations
$[\sigma_{|\psi|}]_\alpha$ of
$|\psi|$ within XY cells of type $\alpha$ from the definition
\begin{equation}
  \left[\sigma_{|\psi|}\right]_\alpha =
  \left[\sqrt{ \frac{\langle|\psi_i|^{2}\rangle-\langle|\psi_i|\rangle^{2}}
  {\langle|\psi_i|\rangle^{2}}  }\right]_\alpha,
  \label{eq:sigma_avg}
\end{equation}
where the triangular brackets denote a thermodynamic average, and
$\left[...\right]_{\alpha}$ denotes a space average over the
$\alpha$ sites. $\left[\sigma_{|\psi|}\right]_\beta$ is computed
analogously. In systems with disorder, the square brackets denote a
disorder average as well as a space average.

Fig.~\ref{fig:ldos_allT} shows the partial LDOS$(\omega,r)$ averaged
over $\alpha$ and $\beta$ cells, at both $t = 0$ and finite $t$. The
systems shown have a fraction $c_{\beta}=0.1$ of $\beta$ sites
randomly distributed. At $t=0$ the $\alpha$ regions show strong,
sharp coherence peaks while the $\beta$ regions have a larger gap
but lower and broader peaks. When the temperature is increased to
$t=0.015$, the heights of both peaks are reduced, and their widths
are increased, but the $\alpha$ peak is still quite sharp, because the
system still has phase coherence.   This temperature is still below
the phase ordering temperature of $t_{c}\simeq 0.03$, as discussed
below.  As $t$ is increased still further, to $t = 0.035$ and $t =
0.055$, the two density of states peaks broaden still further, there
is scarcely any residue of a gap in the density of states, and there
is now no sign of a real coherence peak in either the $\alpha$ or
the $\beta$ regions.

\begin{figure}[t]\centering
  {\includegraphics[width=10cm]{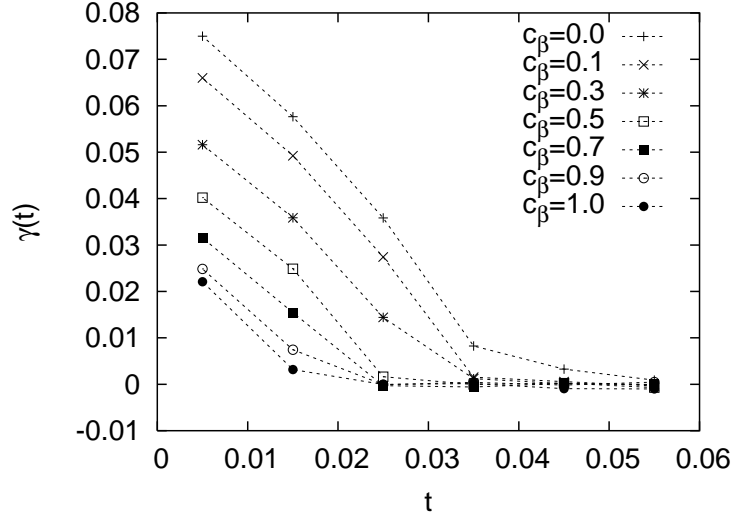}}
  \caption{
    Superfluid density $\gamma(t)$ versus temperature $t$, for systems
    with different concentrations $c_{\beta}$ of $\beta$ cells distributed
    randomly over the atomic lattice. $\beta$ cells have
    $t_{c0} = 0.42$, whereas $\alpha$ cells have $t_{c0} = 0.14$,  However, the
    coupling constant between two nearest neighbor-cells $\langle i j
    \rangle$ includes, at low $t$,
     a factor $1/\sqrt{t_{c0i}\,t_{c0j}}$, which
    results in a suppression of the superfluid density in systems with
    large concentrations of $\beta$ cells [see Eq.~(\ref{eq:JXY3})].}
  \label{fig:hel}
\end{figure}

In Fig.\ \ref{fig:hel}, we show the superfluid density $\gamma(t)$,
for the model just described but for various concentrations
$c_\beta$ of the (randomly distributed) $\beta$ cells.  For $c_\beta
= 0.1$, the phase-ordering transition temperature $t_c \sim 0.03$ in
these units.  Thus, of the plots in the previous Figure,
two are below and two are above the
phase-ordering transition.

\begin{figure}[t]\centering
  {\includegraphics[width=10cm]{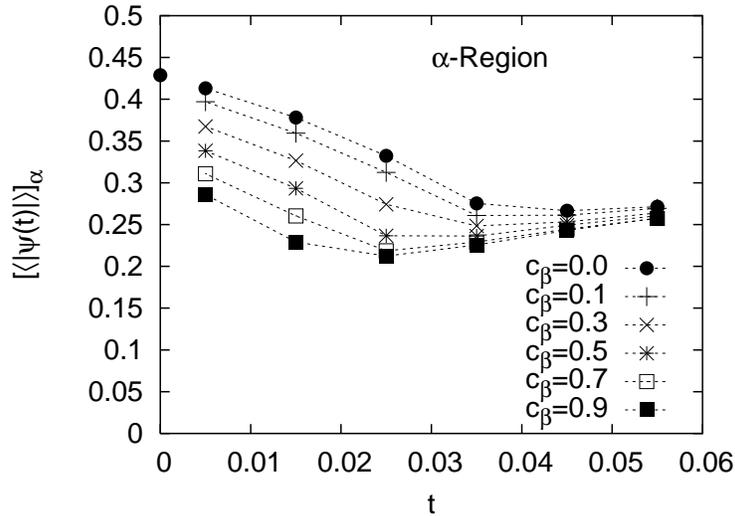}}
  \caption{
    Space, thermally, and disorder-averaged $\left[\langle|\psi|\rangle\right]_\alpha$,
    averaged over $\alpha$ cells, for systems with
    different concentrations $c_{\beta}$ of $\beta$ cells, as described in
    the caption of Fig.~\ref{fig:hel}.  In an $\alpha$ cell
    $t_{c0}=0.14$ while $t_{c0}=0.42$ in a $\beta$ cell.}
  \label{fig:psi_abs_a}
\end{figure}

\begin{figure}[t]\centering
  {\includegraphics[width=10cm]{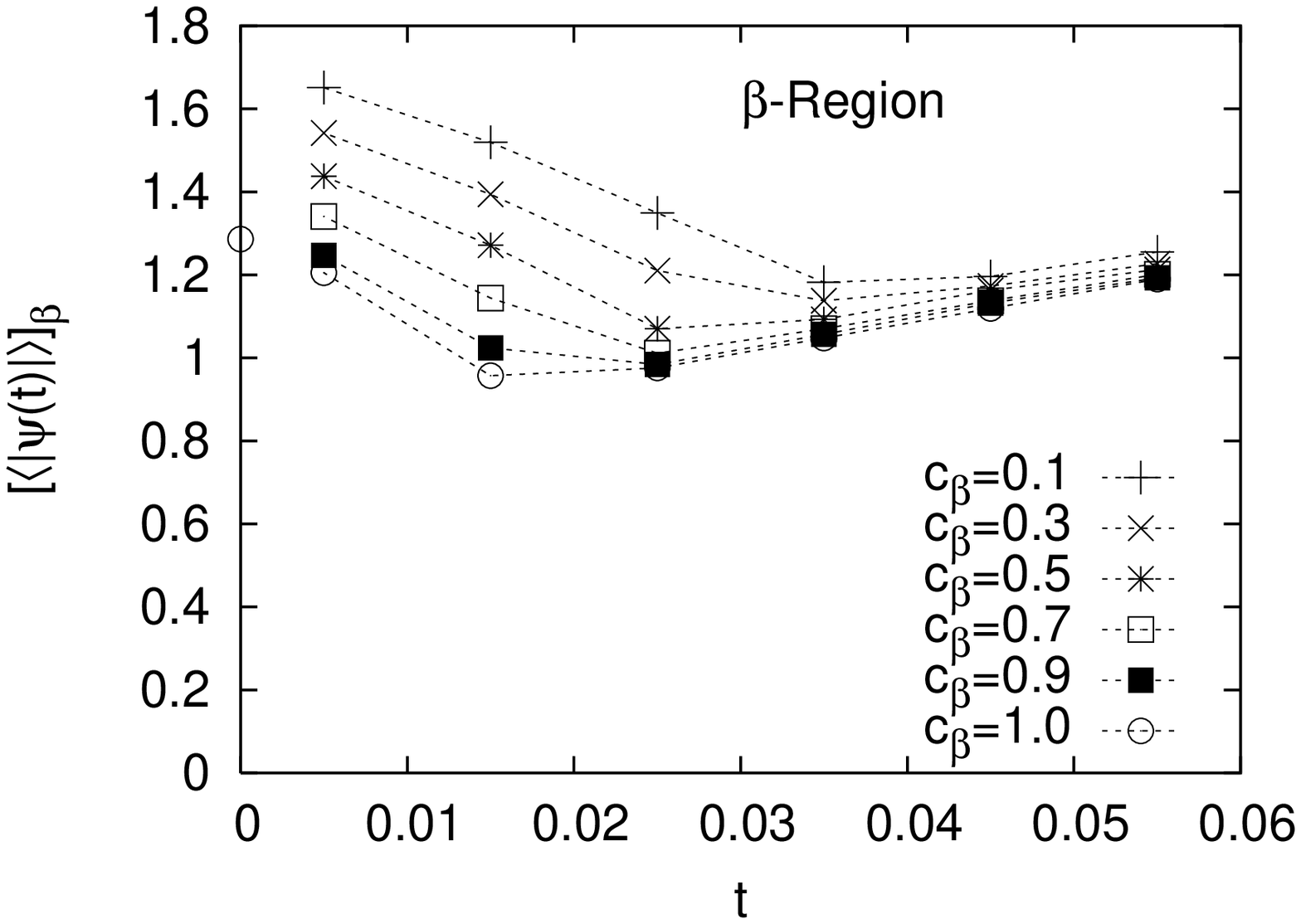}}
  \caption{
    Same as Fig.~\ref{fig:psi_abs_a}, but averaged over
    the $\beta$ cells.}
    \label{fig:psi_abs_b}
\end{figure}

\begin{figure}[t]\centering
  {\includegraphics[width=10cm]{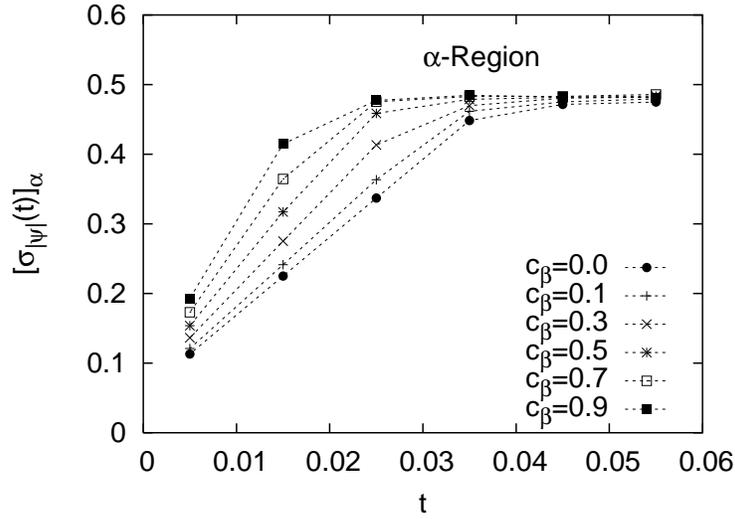}}
  \caption{
    Relative thermal fluctuations $[\sigma_{|\psi|}(t)]_\alpha$ of $|\psi|$, averaged over the
    $\alpha$ cells, for
    systems with different concentrations $c_{\beta}$ of $\beta$
    as shown in Fig.~\ref{fig:hel}.}
  \label{fig:rel_fluct_a}
\end{figure}

\begin{figure}[t]\centering
  {\includegraphics[width=10cm]{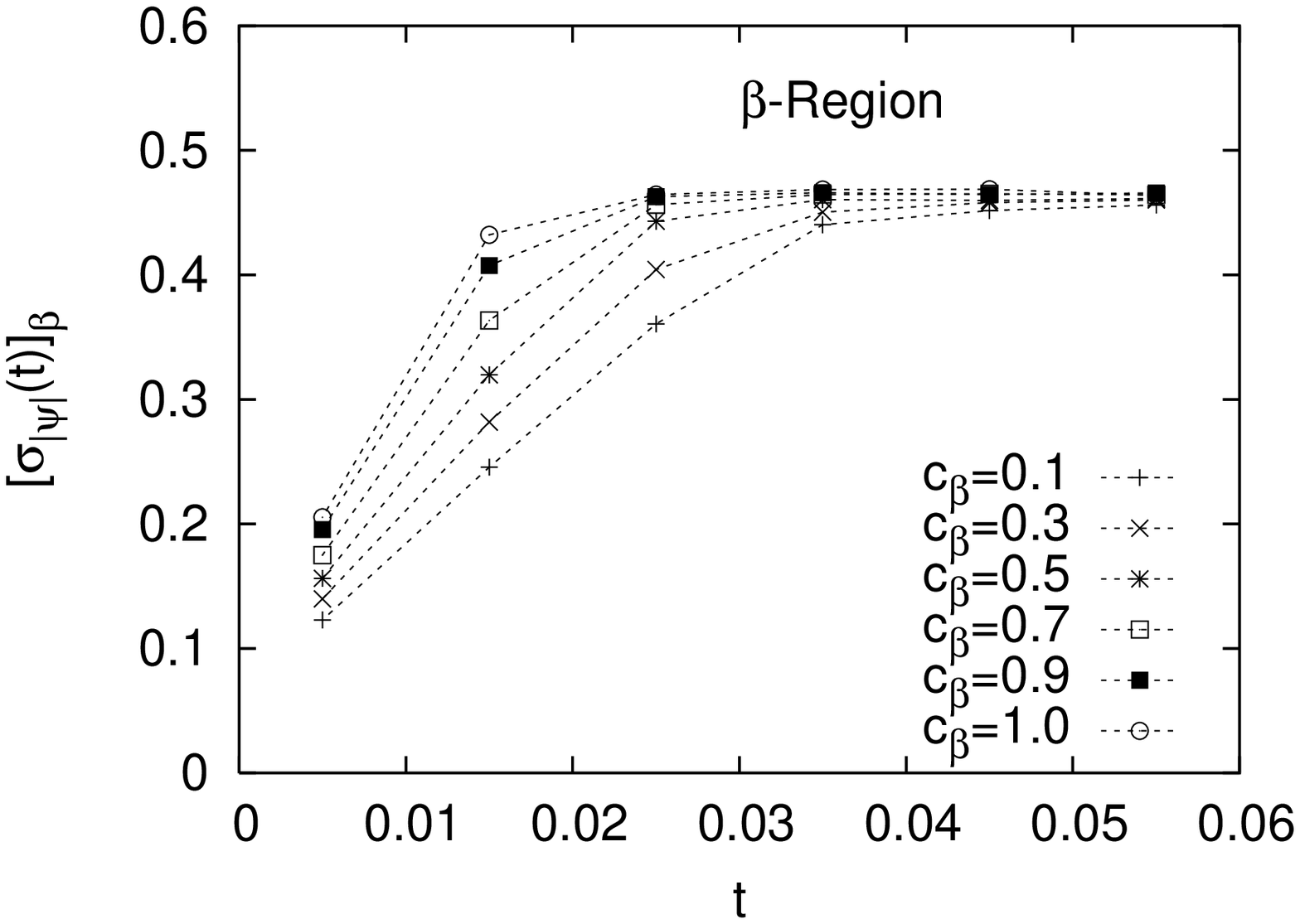}}
  \caption{
    Same as Fig.~\ref{fig:rel_fluct_a} but averaged over the $\beta$ cells.}
  \label{fig:rel_fluct_b}
\end{figure}

In Figs.\ \ref{fig:psi_abs_a} and \ref{fig:psi_abs_b}, we show the
thermal, spatial, and disorder averages of $|\psi|$ over the
$\alpha$ and $\beta$ regions, denoted $[\langle|\psi|\rangle]_\alpha$
and $[\langle|\psi|\rangle]_\beta$,
while Figs.\ \ref{fig:rel_fluct_a},
and \ref{fig:rel_fluct_b} show the corresponding averages of the
root-mean-square fluctuations $\sigma_{|\psi|}$.  A number of
features deserve mention.  First, the average $|\psi|$ is, of
course, larger in the $\beta$ regions than in the $\alpha$ regions,
but the root-mean-square fluctuations are comparable in each of the
two regions.  Second, the increases in the averages of $|\psi|$
above the phase-ordering temperature is an artifact of a
Ginzburg-Landau free energy functional as we now explain in detail.
%
%

The asymptotic behavior of $|\psi|^2$ as $t\rightarrow \infty$, for
homogeneous systems, can be obtained in the following way : At very
high temperatures $t \gg t_{c0}$, the first term
in~(\ref{eq:GLgeneral4}) goes like $\sim|\psi|^2$, the second goes
like $\sim|\psi|^4/t$ and the third (coupling) term goes like
$\sim|\psi|^2/t$. We can then neglect the contribution of the third
term, whence at large $t$ the XY cells are effectively decoupled.
The thermal average of $|\psi|^2$ for an isolated cell is given by
\begin{equation}
  \langle|\psi|^2\rangle = \frac{ \int_{0}^{\infty}|\psi|d|\psi| |\psi|^2  \exp(-f|\psi|^2-g|\psi|^4) }
  {  \int_{0}^{\infty}|\psi|d|\psi|  \exp(-f|\psi|^2-g|\psi|^4) }.
  \label{eq:isolated1}
\end{equation}
In our case,
\begin{equation}
  f=\frac{K_{1}}{t_{c0}^{3}\lambda^{2}(0)E_{0} }
  \label{eq:f}
\end{equation}
and
\begin{equation}
  g=\frac{K_{1}}{2\,(9.38) t_{c0}^{4}\lambda^{2}(0)E_{0} t}.
  \label{eq:g}
\end{equation}
If $f$ and $g$ are real and positive, as in the present case, the
integrals appearing in (\ref{eq:isolated1}) can be carried out, with
the result
\begin{equation}
  \langle|\psi|^2\rangle = -\frac{f}{2g} +
  \frac{\exp(-f^2/4g)}{\sqrt g \pi \text{erfc}\,\,(f/2\sqrt{g})}.
  \label{eq:isolated2}
\end{equation}
Here $\text{erfc}(z)=1-\text{erf}(z)$, $\text{erf}(z)$ being the
gaussian error function.  Using an asymptotic
expansion~\cite{arfken} for $\text{erf}(f/2\sqrt{g}$), applicable
when $f/\sqrt{g}>>1$ as in the present case, we can show that
\begin{equation}
  \lim_{t\rightarrow\infty}\langle|\psi|^2\rangle \rightarrow 1 / f.
  \label{eq:asymptotic}
\end{equation}
Substituting $f$ from eq.\ (\ref{eq:f}) leads to
\begin{equation}
  \lim_{t\rightarrow\infty}\langle|\psi|^2\rangle \rightarrow \frac{t_{c0}^{3}\lambda^{2}(0)E_{0} }{K_{1}} \simeq 0.6
  \label{high_t_psi2}
\end{equation}
where the last approximate equality is obtained using the parameters
we have discussed above, namely $K_{1}\simeq 2866$ eV-$\AA^2$,
$\lambda(0)$ = 1800 $\AA$, $t_{c0}=0.14$, $E_{0}=200$meV.

On the other hand, using eq.\ (\ref{eq:psi}), we obtain
\begin{equation}
  \lim_{t\rightarrow\ 0}\langle|\psi|^2\rangle  \simeq 0.2
  \label{low_t_psi2}
\end{equation}
Thus, our model introduces an unphysical finite value of
$\langle|\psi|^2\rangle$ at large $t$. This behavior has been
observed in other studies of similar models\cite{bittner_janke},
while in other investigations this feature is less obvious because
of the parameters used\cite{ebner_stroud25}. In our case, since we
are interested in temperatures $t < t_{c} < t_{c0}$ or $t\sim
t_{c}<t_{c0}$, this unphysical high-temperature behavior should not
be relevant to our calculations.


Our results for $\left[\langle |\psi| \rangle\right]_\alpha $ and
$\left[\langle |\psi|\rangle\right]_\beta$ suggest an explanation
for one feature in the plots of the LDOS (see Fig.\ 7). Namely, the
$\beta$ peak generally occurs at {\em higher} $\omega$ than it would
in a superconductor made entirely of $\beta$ material.   This shift
occurs because, when the $\beta$ and $\alpha$ regions are mixed,
$\left[\langle|\psi|\rangle\right]_\beta$ is larger than its value
in a homogeneous $\beta$ system (as we further discuss below). This
behavior of $\left[\langle|\psi|\rangle\right]_\beta$ can be seen in
Fig.~\ref{fig:psi_abs_b}, where this quantity is plotted for
different values of $c_{\beta}$.  Clearly, at low $t$, $|\psi|$
increases as $c_{\beta}$ decreases.  For the homogeneous $\beta$
system, $|\psi(t=0)| = 1.29$, as can be obtained directly from eq.\
(\ref{eq:psi}); this value is shown as an open circle at $t = 0$.
This upward shift in the $\left[\langle|\psi|\rangle\right]_\beta$
would be difficult to measure, since a pure $\beta$ material may not
exist.

The behavior of $\left[\langle|\psi|\rangle\right]_\beta$ has an
analog, in our model, in the corresponding behavior in the $\alpha$
cells.  Specifically, if $\alpha$ cells are the minority component
in $\beta$ host, $\langle|\psi|\rangle$ tends to be substantially
smaller than in pure $\alpha$ systems: the smaller the concentration
$c_\alpha = 1 - c_\beta$, the smaller the value of $\langle
|\psi|\rangle$ in those regions [see Fig.\ \ref{fig:psi_abs_a}]. The
behavior of $\langle |\psi|\rangle$ in both $\alpha$ and $\beta$
regions basically follows from our earlier discussion, according to
which $|\psi|^2$ is larger in regions with a small gap.

\section{Discussion}


We have presented a phenomenological model for the
temperature-dependent single-particle density of states in a BCS
superconductor with a $d_{x^2-y^2}$ order parameter.  Our model
includes both inhomogeneities in the gap magnitude and fluctuations
in the phase and amplitude of the gap. While some of these features
have been included in previous models for the density of states (e.
g. phase fluctuations in a homogeneous d-wave superconductor,
inhomogeneities in the gap magnitude at $T = 0$), our model is more
general, and thus potentially more realistic for some cuprate
superconductors.

Our main goal is to examine the properties of an inhomogeneous
superconductor, including many effects which are likely to be
significant in real cuprate materials.  The amplitude and phase
fluctuations are treated by a discretized Ginzburg-Landau free
energy functional, while the density of states is obtained by
solving the Bogoliubov-de Gennes equations for a superconductor with
a tight-binding density of states and a $d_{x^2-y^2}$ energy gap.

In all our calculations, we have assumed that the superconductor has
two types of regions: $\alpha$, with a small gap but a high
superfluid density; and $\beta$, with a large gap and small
superfluid density. This assumption appears consistent with many
experiments on the high-T$_c$ cuprates, especially in the underdoped
regime \cite{lang_davis}.
If we assume that the minority component is of type $\beta$,
embedded in an $\alpha$ host, we find that the local density of
states at $T = 0$ at the $\alpha$ sites has sharp coherence peaks,
whereas that of the $\beta$ sites is substantially broadened. This
behavior is similar to experiment \cite{lang_davis, pan_davis}.

This description applies to a {\em disordered} distribution of
$\beta$ sites in an $\alpha$ host. If the $\beta$ sites are,
instead, arranged on a lattice, the local density of states on the
$\beta$ sites is sharper, but also has distinct oscillations as a
function of energy.  Since such oscillations are absent in
experiments, the actual $\beta$ regions, if they exist as a minority
component, are probably distributed randomly.

In the reverse case of $\alpha$ regions embedded randomly in a
$\beta$ host, neither component has an extremely sharp density of
states peak.   While the $\alpha$ peak is still quite sharp, it is
broader than the $\alpha$ peak in the $\beta$-minority case.  This
result suggests that, if one component occurs only as isolated
regions, its minority status tends to broaden its coherence peaks.

Our results also show that the local density of states is strongly
affected by phase fluctuations.  This feature has already been found
for a {\em homogeneous} d-wave superconductor\cite{eckl}, but here
we demonstrate it in an inhomogeneous superconductor.   The most
striking effect of finite $T$ is that the coherence peak in the
$\alpha$ component disappears above the phase-ordering transition
temperature $T_c$.  The $\beta$ component does not show a coherence
peak even at very low temperatures, but nonetheless this peak too is
significantly broadened above $T_c$.   For $T$ well above $T_c$,
there is no appreciable gap in the local density of states either at
the $\alpha$ or the $\beta$ sites.

Our calculations include thermal fluctuations in the amplitude as
well as the phase of $\psi$.  In general, thermal amplitude
fluctuations seem to have only a minor influence on the local
density of states.  By contrast, the variations in $|\psi|$ due to
{\em quenched} disorder (i. e., the presence of $\alpha$ and $\beta$
regions in our model) strongly affect the local density of states,
as we have already described.  To check on the influence of purely
thermal amplitude fluctuations, we have calculated the density of
states of a homogeneous $\alpha$ superconductor with both phase and
amplitude fluctuations, and have compared this to a similar
calculation with only phase fluctuations. We find that the
additional presence of amplitude fluctuations has little effect on
the density of states.
%

To smooth the local density of states, we include in our density of
states calculations a magnetic field equal to a flux $hc/e$ in the
entire sample area, following the method of Assaad\cite{assaad}.
This field greatly smooths the local density of states, which
otherwise varies extremely sharply with energy, because of the many
degenerate states of a finite sample at zero field.   Our calculated
density of states does, of course, correspond to a physical magnetic
field, and thus differs slightly from that at zero field.  For
example, in a homogeneous system with a finite $d$-wave gap, the
LDOS$(\omega)$ goes to zero as $|\omega| \rightarrow
0$.
By contrast, at finite  field, the LDOS approaches a constant value
at low $|\omega|$.   With no gap, our calculated DOS with nonzero
field is indistinguishable from that of a conventional 2D
tight-binding band (see Fig.\ 1), because the field is low
(typically around 0.002 flux quanta per atomic unit cell).  We
conclude that the weak magnetic field very effectively smooths the
calculated LDOS in a finite two-dimensional sample with a d-wave
gap, but produces a density of states similar to that at zero field,
except at very low $|\omega|$.

Although we have included this magnetic field in the Bogoliubov-de Gennes equations,
we have omitted it from the Ginzburg-Landau free energy functional, which is thus
that of a zero-field system. As we now discuss, we believe that this numerical scheme should
indeed converge to the correct physical result for zero magnetic field in the limit
of a large computational sample.

As noted earlier, we introduce the vector potential into the LDOS calculation in order to smooth the 
resulting density of states.  In the limit of a large system, the effect of the vector potential, 
corresponding to a single quantum of flux, should become negligible, since the flux density becomes
very small.  This is already suggested by our calculated results for the two system sizes we consider (see
Fig.\ 2 and the corresponding discussion).
Even with a finite superconducting gap, the vector potential affects the LDOS very little, except at low energies; moreover, even this effect becomes smaller as the sample size increases.  Therefore, in the limit of a large
enough sample, our approach should give a very similar LDOS to one calculated with no vector potential.
Hence, it is reasonable to use this approach in combination with a zero-field Ginzburg-Landau free
energy functional to calculate the LDOS at finite temperatures.

If we were to introduce a similar field into the Ginzburg-Landau functional, we believe that
it would have a substantial effect, not due to smoothing, on the phase ordering.  There would be, not only
the XY-like phase transition, as at zero field, but also additional phase fluctuations
arising from the extra field-induced vortex.  Since this extra vortex is absent at zero
field, these effects would be irrelevant to the zero-field system we wish to model.
By contrast, introducing a field into the Bogoliubov-de Gennes equations, as we do,
provides desirable smoothing with little change in the calculated LDOS; moreover, even this slight
change decreases with increasing sample size.  Therefore, we believe that
the best way to obtain a smooth LDOS at both zero and finite temperatures is to introduce the
vector potential into the Bogoliubov-de Gennes equations, for smoothing purposes, but not to 
include it in the Ginzburg-Landau free energy functional.  Our numerical results suggest that this
procedure is indeed justified.  

One feature of our numerical results may seem counterintuitive. In
our model, the $\alpha$ component is assumed to have a gap three
times smaller than that of $\beta$, but has a larger local
superfluid density, i. e., a smaller penetration depth.   We then
find that the gap in the $\alpha$ region is {\em smaller} in a
two-component system with both $\alpha$ and $\beta$ regions, than it
is in a pure $\alpha$ system. This counterintuitive result, however,
emerges naturally from our discrete Ginzburg-Landau model, which is
minimized if the quantities $\Delta_i/(\lambda_i(0)T_{c0i})$ are
equal.  For our model, $\lambda_i(0)$ is smaller in the small-gap
material.  It would be of interest if experimental evidence of this
behavior were found in a real material.

In our calculation, the LDOS is obtained from a Bogoliubov-de Gennes
Hamiltonian whose parameters are determined by the Ginzburg-Landau
functional.   In fact, it should be possible to proceed in the
opposite direction, and obtain the parameters of the functional from
the LDOS.  Specifically, the energy required to change a phase
difference by a given amount depends on an integral over the LDOS.
Thus, the calculation we have presented in this paper can, in
principle, be made fully self-consistent.

\section{Acknowledgments}  We are grateful for support through the
National Science Foundation grant DMR04-13395.  The computations
described here were carried out through a grant of computing time
from the Ohio Supercomputer Center.   We thank E. Carlson, S.
Kivelson, M. Randeria, N. Trivedi, R. Sensarma, D. Tanaskovic, and K. Kobayashi
for valuable discussions.

\newpage


\begin{figure}[ht]\centering
  {\includegraphics[width=10cm]{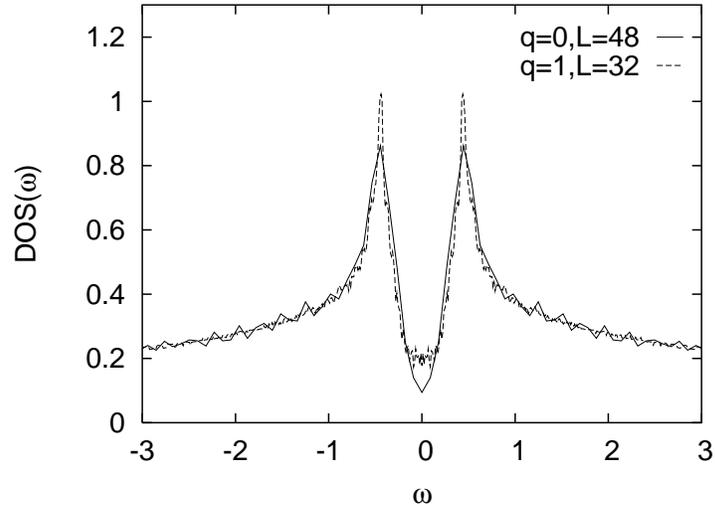}}
  \caption{
    Comparison of the
    $T=0$ DOS of a small system ($32
    \times 32$ atomic lattice)
    containing one quantum of magnetic field ($q=1$) to that of a larger ($48 \times
    48$) system with no magnetic field ($q=0$).  Both systems are
    homogeneous with $t_{c0}=0.14$. Except at very low energies,
    the magnetic field produces little
    change in the shape of the curve.}
    \label{fig:ldos_flux0_flux1}
\end{figure}



\end{document}